\documentclass[apj,numberedappendix,appendixfloats]{emulateapj}
\usepackage{amsmath}
\usepackage{graphicx}
\usepackage{booktabs,threeparttable}
\usepackage{hyperref}
\usepackage{multirow}
\usepackage{float}
\usepackage{ulem}
\shorttitle{Geometry, Kinematics, and Magnetization of Simulated Prestellar Cores}
\shortauthors{Chen \& Ostriker}

\begin{document}

\title{Geometry, Kinematics, and Magnetization of Simulated Prestellar Cores}
\author{Che-Yu Chen\altaffilmark{1} and Eve C. Ostriker\altaffilmark{2}}
\email{cc6pg@virginia.edu}
\altaffiltext{1}{Department of Astronomy, University of Virginia, Charlottesville, VA 22904, USA}
%\altaffiltext{2}{Department of Astronomy, University of Maryland, College Park, MD 20742, USA}
\altaffiltext{2}{Department of Astrophysical Sciences, Princeton University, Princeton, NJ 08544, USA}

\begin{abstract}

We utilize the more than 100 gravitationally-bound dense cores formed in our three-dimensional, turbulent MHD simulations reported in \cite{2015ApJ...810..126C} to analyze structural, kinematic, and magnetic properties of prestellar cores.
Our statistical results disagree with the classical theory of star formation, in which cores evolve to be oblate with magnetic field parallel to the minor axes.
Instead, we find that cores are generally triaxial, although the core-scale magnetic field is still preferentially most parallel to the core's minor axis and most perpendicular to the major axis.  The internal and external magnetic field directions are correlated, but the direction of integrated core angular momentum is misaligned with the core's magnetic field, consistent with recent observations.  The ratio of rotational/total kinetic and rotational/gravitational energies are independent of core size and consistent in magnitude with observations.  The specific angular momentum also follows the observed relationship $L/M \propto R^{3/2}$, indicating rotation is acquired from ambient turbulence.  With typical $E_\mathrm{rot}/E_K \sim 0.1$, 
rotation is not the dominant motion when cores collapse.

\end{abstract}


\section{Introduction}
\label{sec:intro}

In molecular clouds (hereafter MCs), multi-scale supersonic flows compress material and initiate creation of filamentary structures \citep{2014prpl.conf...27A}. Within filaments, some of the overdense regions will shrink under self-gravity to form prestellar cores and then collapse to create protostellar systems, which later become stars \citep{1987ARA&A..25...23S}. 
Dense cores are therefore the immediate precursors of at least low-mass stars or close binary systems \citep{2007ARA&A..45..565M}. Their properties provide the initial conditions of star formation, and determine the local environment of protostellar disks and outflows.

Cores are observed in dust continuum and molecular lines.
Recent results from the Herschel Gould Belt Survey \citep{2010A&A...518L.102A} suggest that dense cores are mostly associated with filaments (\citealt{2010A&A...518L.106K}; or see review in \citealt{2014prpl.conf...27A}). This association is consistent with the theoretical expectation that thermally supercritical filaments (mass-per-unit-length $M/L > 2 {c_s}^2/G$; \citealt{1964ApJ...140.1056O}) would fragment longitudinally into cores \citep[e.g.][]{1992ApJ...388..392I,1997ApJ...480..681I}.
However, 
in a turbulent environment like a MC, dense filaments are not quiescent structures in which perturbations slowly grow. Rather,
various simulations with turbulence have shown that filaments and cores develop simultaneously \citep[e.g.][]{2011ApJ...729..120G,2014ApJ...785...69C,2015ApJ...810..126C,2014ApJ...791..124G,2014ApJ...789...37V,2015ApJ...806...31G}, in contrast to the two-step scenario, because multi-scale growth is enabled by the non-linear perturbation generated by turbulence.

In combination with turbulence and gas gravity, magnetic effects are considered one of the key agents affecting the dynamics of star formation in MCs, at all physical scales and throughout different evolutionary stages \citep{2007ARA&A..45..565M}. 
At earlier stages and on larger scales, the magnetic field can limit compression in turbulence-generated interstellar shocks that create dense clumps and filaments \citep{1956MNRAS.116..503M}. 
Meanwhile, the core-scale magnetic field is expected to be important in affecting the gas dynamics within cores, and is interconnected with the cloud-scale magnetic field. 
The magnetic field within collapsing cores provides the main channel for the gas to lose angular momentum via ``magnetic braking" during the collapse of prestellar cores and the formation of protostellar disks (\citealt{1956MNRAS.116..503M,1974Ap&SS..27..167G,1976ApJ...207..141M,1991ApJ...373..169M}; see review in \citealt{2014prpl.conf..173L}). 

In strict ideal MHD, magnetic braking can be simply understood as the inner, faster-rotating material being slowed down by the outer, more slowly-rotating material because they are interconnected by magnetic field lines \citep{1979ApJ...230..204M,1980ApJ...237..877M}. Numerical simulations also showed that the formation of rotationally supported disks is suppressed by catastrophic magnetic braking, unless the dense cores are weakly magnetized to an unrealistic level \citep{2003ApJ...599..363A,2008A&A...477....9H,2008ApJ...681.1356M,2011A&A...528A..72H}. 
Many solutions have been proposed to solve this problem, including non-ideal MHD effects \citep{2010ApJ...716.1541K,2011ApJ...738..180L,2011PASJ...63..555M,2012A&A...541A..35D,2013ApJ...763....6T}, turbulence-induced diffusion \citep{2012ApJ...747...21S,2012MNRAS.423L..40S,2013MNRAS.432.3320S,2013A&A...554A..17J}, and the magnetic field-rotation misalignment \citep{2009A&A...506L..29H,2010MNRAS.409L..39C,2012A&A...543A.128J,2013ApJ...767L..11K}. 
The initial magnetic field structure and strength within prestellar cores are therefore important for late evolution during core collapse, since they control the efficiency of this magnetic braking process.

Rotation is also important in the evolution leading to the creation
of protostellar systems within dense cores. The angular momentum of star-forming cores is a critical parameter in protostellar evolution, but its origin is not well understood \citep[see review in][]{2014prpl.conf..173L}. 
It is known that some dense cores show a clear gradient in line-of-sight velocity, while others have a relatively random velocity field \citep[e.g.][]{1993ApJ...406..528G,2002ApJ...572..238C}. The observed velocity gradient is commonly used, when present, to estimate a core's angular momentum.
Observational and theoretical understanding of core angular momentum is important as this property is essential to subsequent evolution:
whether a single star or multiple system is formed \citep{2002ARA&A..40..349T}, and whether a large or small disk is produced \citep[see review in][]{2014prpl.conf..173L}. 

There have been several observational projects aimed at resolving the velocity structure within dense cores. Linear fitting is generally applied to observed velocity gradients across cores, regardless of the complex nature of the velocity field. It is assumed that rigid body rotation applies and that the angular speed is roughly the gradient of line-of-sight velocity \citep{1993ApJ...406..528G}. 
Previous observations \citep{1993ApJ...406..528G,2002ApJ...572..238C,2003A&A...405..639P,2007ApJ...669.1058C,2011ApJ...740...45T} have found a power-law relationship between the specific angular momentum, $L/M$, and radius, $R$, for dense cores/clumps with radii $\sim 0.01-1$~pc, $L/M \propto R^\alpha$ with $\alpha \approx 1.5$. 

The $L/M-R$ correlation over a huge range of spatial scales suggests that gas motion in cores originates at scales much larger than the core size, or the observed rotation-like features may arise from sampling of turbulence at a range of scales \citep{2000ApJ...543..822B}. 
Simulated cores from our previous work \citep[][hereafter CO14 and CO15]{2014ApJ...785...69C,2015ApJ...810..126C} provide a suitable database to test whether the $L/M-R$ correlation extends to even smaller scales ($R$~$\sim$~0.008--0.05~pc), and whether the assumption of rigid body rotation is valid in simulated cores.
These questions are one focus of the present study.

In this paper, we present our results on structural, kinematic, and magnetic properties of simulated prestellar cores formed in three-dimensional (3D) MHD simulations; the simulation ingredients include convergent flow, multi-scale turbulence, and gas self-gravity. This set of simulations was first introduced in \hyperlink{CO15}{CO15} to investigate the forming process of prestellar cores and how it correlates with the cloud environment. We have shown in \hyperlink{CO15}{CO15} that these cores have masses, sizes, and mass-to-magnetic flux ratios similar to the observed ones. Here, we extend our previous study to include analysis of the geometry and kinematic features of cores, as well as the relative direction of core-scale magnetic field. 

The outline of this paper is as follows. We describe our simulation models, numerical methods, and analytic algorithms in Section~\ref{sec::method}. Our results on core geometry and other structural properties are presented in Section~\ref{sec::physical}, while Section~\ref{sec::dyn} focuses on the kinetic features of dense cores. Based on these results, we discuss the origin of core angular momentum in Section~\ref{sec::discussion}. Finally, Section~\ref{sec::summary} summarizes our conclusions.

\section{Method}
\label{sec::method}

\begin{figure*}
\begin{center}
\includegraphics[width=\textwidth]{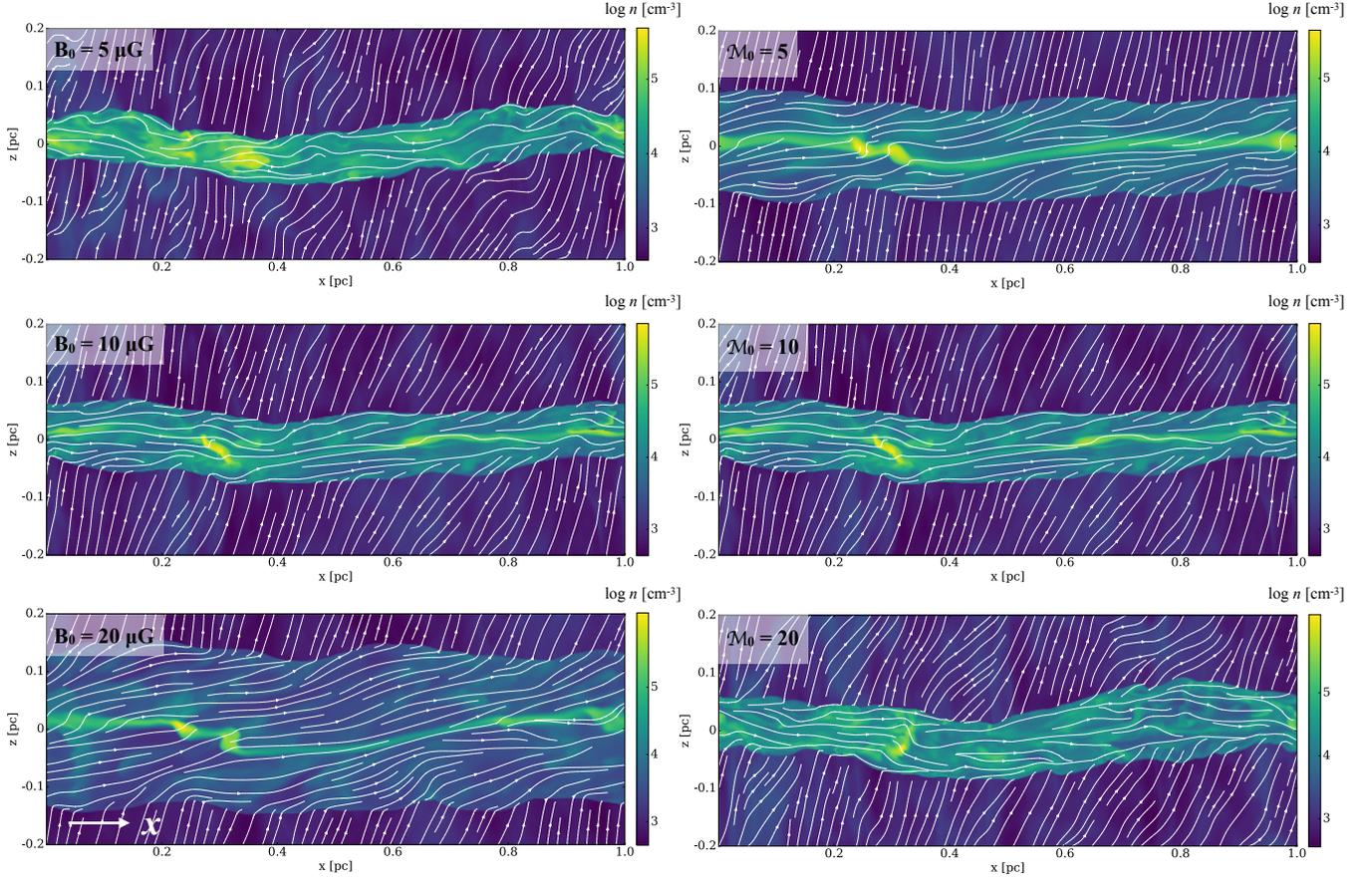}
\caption{The density ({\it colormap}) and magnetic field ({\it white streaklines}) in slices cut through the center of the simulation box at $y=0.5$~pc, for each model considered in this study. The central sub-layer, which is a consequence of secondary convergent flow \citep[see][]{2017ApJ...847...140C}, can be clearly seen for models B20 and M5 ({\it bottom-left} and {\it top-right} panels), while in models B5 and M20 ({\it top-left} and {\it bottom-right} panels) the entire post-shock layers remain homogeneous. 
}
\label{pslayer}
\end{center}
\end{figure*}

\subsection{Numerical Simulations}

The simulations we analyze here are described in \hyperlink{CO14}{CO14}, \hyperlink{CO15}{CO15} and summarized here. 
These isothermal, self-gravitating ideal-MHD simulations consider the immediate surroundings of strongly-magnetized core-forming regions ($1$~pc$^3$) within MCs. 
For all simulations, the numerical resolution is $512^3$, yielding cells of individual size $\Delta x \approx 0.002$~pc.
The model establishes super-Alfv{\'e}nic convergent flows (powered by the cloud-scale supersonic turbulence) compressing diffuse, turbulent gas to form denser, star-forming clumps in post-shock regions (see e.g.~Figure~\ref{pslayer}). 
The idealized setup of a local converging turbulent flow gives us better control of the post-shock environment based on the simulation initial conditions as discussed in \hyperlink{CO14}{CO14}, \hyperlink{CO15}{CO15}, and \cite{2017ApJ...847...140C}, which is crucial for analyzing the connection between the properties of prestellar cores and the environment they formed within.

The main model parameters (inflow Mach number ${\cal M}_0 \equiv v_{0,z}/c_s$ and background magnetic field strength $B_0$) are listed in Table~\ref{CoreSum}. 
For all simulations, we initialize the background magnetic field in the $x-z$ plane with an angle $20^\circ$ with respect to the converging flow along $\pm\hat z$.
We consider two sets of parameters (M5, M10, M20 for varying ${\cal M}_0$ and B5, B10, B20 for varying $B_0$, where M10 and B10 are the same model), which generates a range of core-forming environment and density structures in the post-shock region (Figure~\ref{pslayer}; also see Figure~3 of \hyperlink{CO15}{CO15}).
The post-shock conditions are determined by a combination of many mechanisms including the jump conditions of oblique MHD shocks (see \hyperlink{CO14}{CO14} and \hyperlink{CO15}{CO15}) and the dynamics of secondary convergent flow \citep{2017ApJ...847...140C}. 

Though we list the averaged post-shock plasma beta $\beta_\mathrm{ps} \equiv 8\pi\rho_{\rm ps}{c_s}^2/{B_{\rm ps}}^2$ and Alfv{\'e}n Mach number ${\cal M}_\mathrm{A, ps} \equiv v_{\rm ps}/v_{\rm A, ps} = v_{\rm ps}\cdot \sqrt{4\pi\rho_{\rm ps}} / B_{\rm ps}$ for each simulation model in Table~\ref{CoreSum}, we note that one should not treat the post-shock environment using one single value of these parameters. 
This is especially true for those conditions that generate a dense sub-layer at the mid-plane of the post-shock region.
Figure~\ref{pslayer} shows the density (in $\log$~cm$^{-3}$) and magnetic field directions (white streaklines) in slices cut through the center of each simulation box at $y=0.5$~pc. 
The large-scale converging flow is along the $z$ axis.
A dense, thin sub-layer can be clearly seen in models B20 and M5 but not in models B5 and M20, which generally agrees with the prediction in \cite{2017ApJ...847...140C}. 
This cannot be derived from the averaged post-shock conditions ($\beta_\mathrm{ps}$ and ${\cal M}_\mathrm{A, ps}$) listed in Table~\ref{CoreSum}.
As discussed in \cite{2017ApJ...847...140C}, this sub-layer is created by supersonic secondary convergent flows and thus is relatively stagnant, which means dense clumps within it will undergo a more quiescent process to form prestellar cores. Roughly speaking, we expect the core-forming environment created in models B20 and M5 to be more dominated by the magnetic field, while gas turbulence plays a more critical role during core formation in models B5 and M20. We therefore rearrange the order of the models to group B20 and M5 as well as B5 and M20 in our following analysis and plots.

For each set of model parameters, we run 6 simulations with different realizations of the input turbulence. At the time the most evolved core collapses (when $n_{\rm max} \geq 10^7~{\rm cm}^{-3}$), we identify gravitationally bound cores as regions with $E_{\rm grav} + E_{\rm thermal} + E_{\rm mag} < 0$ (see \hyperlink{CO14}{CO14}, \hyperlink{CO15}{CO15} and below for details).  This yields a total of 186 well-resolved self-gravitating cores within the post-shock layer.

\subsection{Measuring Core Properties}
\label{sec::grid}

\begin{figure}
\begin{center}
\includegraphics[width=\columnwidth]{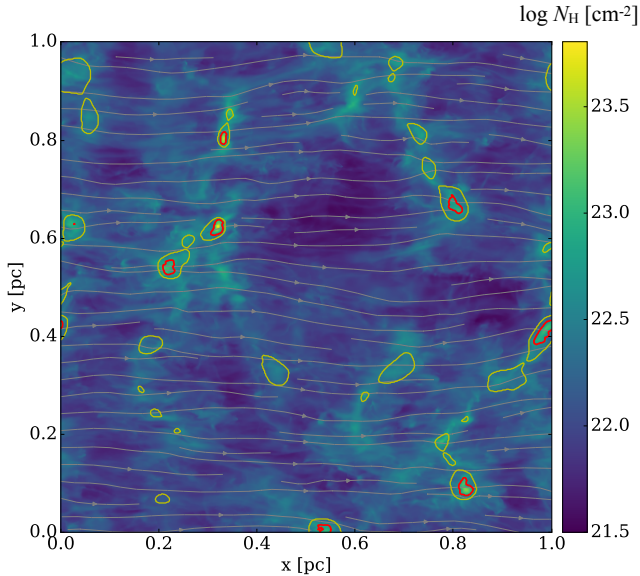}
\caption{Sample map from one of the core-forming simulations considered in this study, from model B5 and realization 1. While the {\it GRID} core-finding routine defines cores in 3D space, here we show the projection of identified cores along the $z$-direction, overlapped on integrated column density of the gas ({\it colormap}) and density-weighted mean magnetic field ({\it grey streaklines}). Both gravitationally unbound ({\it yellow contours}) and bound ({\it red contours}) cores are shown in this map; however we only consider gravitationally bound cores in this study.}
\label{corefind}
\end{center}
\end{figure}

For each simulation, we apply the {\it GRID} core-finding method to identify dense cores using the largest closed gravitational potential contours around single local potential minimums. The original {\it GRID} core-finding routine is developed and discussed in \cite{2011ApJ...729..120G}, while the MHD extension (to include measurement of cores' mass-to-magnetic flux ratios) is implemented by and adopted in \hyperlink{CO14}{CO14} and \hyperlink{CO15}{CO15}. Here, we further update the {\it GRID} core-finding method with angular momentum evaluation as well as derivation of rotational and turbulent energies.
In this study, we also investigate the geometry of dense cores by applying the principal component analysis (PCA) to define the three axes of the core ($a$, $b$, and $c$ from longest to shortest), and to calculate the aspect ratios \citep[see e.g.][]{2011ApJ...729..120G,2015ApJ...806...31G}.
Note that, similar to our previous studies (\hyperlink{CO14}{CO14}, \hyperlink{CO15}{CO15}), identified cores with less than $27$ cells are not considered in the analysis. 
Figure~\ref{corefind} is a sample map showing the simulated gas structure (in column density integrated along $z$) and examples of identified cores from the {\it GRID} core-finding method.

\subsection{Core Angular Momentum}

For each core, it is straightforward to define and calculate the net angular momentum $\mathbf{L}$ by integrating the relative angular momentum of each cell with respect to the center of mass over the whole volume:
\begin{equation}
\mathbf{L} = \sum_i \rho_i \Delta V \cdot \left(\mathbf{r}_i - \mathbf{r}_\mathrm{CM}\right)\times\mathbf{v}_i
\label{eq::netL}
\end{equation}
($\Delta V = \Delta x^3$ is the volume of a single simulation cell).
This vector, together with the coordinate of the center of mass, determines the principal rotational axis for this core, $\hat {\bf L} = {\bf L}/L$; here $L = |{\bf L}|$ is the magnitude of the net angular momentum of the core.
The total rotational inertia of the core around this rotational axis can be derived by firstly determining the projected radius for each cell:
\begin{equation}
\mathbf{r}_{i,\perp} = \left(\mathbf{r}_i - \mathbf{r}_\mathrm{CM}\right) - \left[\left(\mathbf{r}_i - \mathbf{r}_\mathrm{CM}\right) \cdot \hat{\mathbf{L}}\right]\hat{\mathbf{L}}
\end{equation}
and then integrating over the whole volume:
\begin{equation}
I \equiv \sum_i \rho_i \Delta V \cdot |\mathbf{r}_{i,\perp}|^2
\end{equation}
The mean angular velocity $\Omega$ and the net rotational energy $E_\mathrm{rot}$ of the core are
\begin{align}
\Omega & \equiv L / I,\\
E_\mathrm{rot} & \equiv \frac{1}{2}I \Omega^2. \label{eq::Erot}
\end{align}

To compute the turbulent energy within cores under the assumption of rigid-body rotation, we must first calculate the velocity from rotation for each cell:
\begin{equation}
\mathbf{v}_{\mathrm{rot},i} =  r_{i,\perp} \Omega \left(\hat{\mathbf{L}}\times\hat{\mathbf{r}}_{i,\perp}\right).
\label{eq::vrot}
\end{equation}
The turbulent energy of the core under the assumption of rigid-body rotation is then
\begin{equation}
E_\mathrm{turb} = \frac{1}{2}\sum_i \rho_i \Delta V \cdot \left(\mathbf{v}_i - \mathbf{v}_{\mathrm{rot},i}\right)^2.
\label{eq::Eturb}
\end{equation}
The total kinetic energy is therefore defined as $E_{\rm total} \equiv E_{\rm rot} + E_{\rm turb}$.

\subsection{The Ring-fit of Core Rotation}
\label{sec::anacode}

The simple assumption of rigid-body rotation adopted in Equations~(\ref{eq::vrot}) and (\ref{eq::Eturb}) is not always true.
Here we consider a different approach of measuring the core's rotational motion by binning the core into several ``rings" and allowing individual rings to have different rotational speed.

The principal rotational axis given by $\hat {\bf L}$ and $\mathbf{r}_\mathrm{CM}$ defines a cylindrical coordinate system for the core: $(\mathbf{r}_{i,\perp}, h_i)$, where the ``height" can be calculated as
\begin{equation}
h_i =  \left(\mathbf{r}_i - \mathbf{r}_\mathrm{CM}\right) \cdot \hat{\mathbf{L}}.
\end{equation}
We bin the cells by $r_{i,\perp}$ and $h_i$ to define local rings within the core; each ring is chosen to be 3-cell-wide in both radius and height. Similar to the cell-number requirement we applied to cores, if there are fewer than $27$ cells assigned to a ring, that ring is not included in follow-up calculations. We also exclude cores with less than $4$ rings (2 in both $r$ and $h$ directions) to improve the statistics.

For each ring $(r,h)$, we calculate the net angular momentum (through its center of mass) $\mathbf{L}_\mathrm{ring} (r, h)$, the rotational inertia $I_\mathrm{ring} (r, h)$, the mean angular velocity $\Omega_\mathrm{ring} (r, h)$, and the rotational energy $E_\mathrm{rot,ring} (r, h)$ in the same way as Equations~(\ref{eq::netL})-(\ref{eq::Erot}). The ring-fit mean angular velocity for a core is then defined as the inertia-weighted average among $\Omega_\mathrm{ring} (r, h)$:
\begin{equation}
\Omega_\mathrm{ring} = \frac{ \sum\limits_{(r,h)} I_\mathrm{ring} (r, h) \cdot \Omega_\mathrm{ring} (r, h) }{\sum\limits_{(r,h)} I_\mathrm{ring} (r, h) }.
\end{equation}
The total rotational energy derived from the ring fit is
\begin{equation}
E_\mathrm{rot, ring} = \sum_{(r,h)} E_\mathrm{rot,ring} (r, h) .
\end{equation}
where $E_{\rm rot, ring} (r, h)= (1/2) I_{\rm ring} (r, h) \Omega_{\rm ring}(r, h)^2$.
The turbulent energy from the ring fit $E_\mathrm{turb, ring}$ is obtained by following Equations~(\ref{eq::vrot}) and (\ref{eq::Eturb}), but with $\Omega_\mathrm{ring}(\mathbf{r}_{i,\perp}, h_i)$ instead of $\Omega$ and $\hat{\mathbf{L}}_\mathrm{ring} (\mathbf{r}_{i,\perp}, h_i)$ instead of $\hat{\mathbf{L}}$ in Equation~(\ref{eq::vrot}). 
The total kinetic energy from the ring fit is $E_{\rm total, ring} \equiv E_{\rm rot, ring} + E_{\rm turb, ring}$.
By comparing the ratios $E_\mathrm{turb}/E_\mathrm{total}$ and $E_\mathrm{rot}/E_\mathrm{total}$ to $E_\mathrm{turb,ring}/E_\mathrm{total,ring}$ and $E_\mathrm{rot,ring}/E_\mathrm{total,ring}$, we can infer whether rigid-body rotation is a good approximation for our simulated cores (see Section~\ref{sec::nrbRot}).

Note that, because of the selection criteria, the number of cores for which we performed this ring-fit and energy analysis is less than the number of cores considered in the rigid-body rotation analysis discussed in Section~\ref{sec::grid} (see Tables~\ref{CoreSum} and \ref{CoreKE}). Also, the number of cells considered in each ring-fit might be less than the total number of cells within a core, which makes it inappropriate to directly compare the absolute values of $E_\mathrm{rot}$ and $E_\mathrm{rot, ring}$, $E_\mathrm{turb}$ and $E_\mathrm{turb, ring}$, or $E_\mathrm{total}$ and $E_\mathrm{total, ring}$. We therefore use the relative values, $E_\mathrm{rot}/E_\mathrm{total}$ and $E_\mathrm{rot, ring}/E_\mathrm{total,ring}$ when cross-comparing these two fitting methods (see Table~\ref{CoreKE}).

\section{Core Geometry and Physical Properties}
\label{sec::physical}

\renewcommand{\arraystretch}{1.1}
\begin{table*}[t]
\hspace{-.3in}
%\begin{center}
  \begin{threeparttable}
\caption{Properties of identified cores.$^{\dagger}$ }
\label{CoreSum}
\vspace{.1in}
\begin{tabular}{ l | c c || c c | c | c c c c c | c c c c }
  \hline
  Model & ${\cal M}_0$ & $B_0$ & $\beta_\mathrm{ps}$$^\star$ & ${\cal M}_\mathrm{A, ps}$$^\star$ & \# Cores & $b/a$$^{\ddagger}$ & $c/a$$^{\ddagger}$ & $c/b$$^{\ddagger}$ & $\measuredangle[B, a]$$^{\P}$ &$\measuredangle[B, c]$$^{\P}$ & $L/M$ & $\measuredangle[L,a]$$^{\P}$ &$\measuredangle[L,c]$$^{\P}$ & $\measuredangle[B,L]$$^{\P}$ \\ 
   & & ($\mu$G) & & & considered$^{\S}$ & &  &  & ($^\circ$) & ($^\circ$) & ($10^{-4}$ pc$\cdot$km/s)  & ($^\circ$) & ($^\circ$) & ($^\circ$)\\
  \hline
  M5 & 5 & 10 & 0.25 & 0.78 & 32  & 0.61 & 0.28  & 0.49 & 80 & 18 & 5.26 & 77 & 38 & 44\\
  B20 & 10 & 20 & 0.08 & 0.98 & 55  & 0.63 & 0.30 & 0.48 & 81 & 16& 4.38  & 65 & 47 & 51\\
%\hline
  M10B10 & 10 & 10 & 0.16 & 0.81 & 28 & 0.54 & 0.27 & 0.57 & 75 & 27& 5.71  & 67 & 51 & 49\\
%\hline
  B5 & 10 & 5 & 0.35 & 0.86 & 43 & 0.58 & 0.41 & 0.70 & 70 & 42& 5.56 & 77 & 41 & 45\\
  M20 & 20 & 10 & 0.09 & 0.84 & 28 & 0.57 & 0.33 & 0.61 & 79 & 32& 6.77 & 75 & 52 & 59 \\
  \hline 
\end{tabular}
    \begin{tablenotes}
      \footnotesize
      \item $^\dagger$Columns (7)$-$(15) are median values over all cores for each parameter set (6 simulation runs).
      \item $^\star$The post-shock Alfv{\'e}n Mach number is calculated at $t=0.2$~Myr in each model, averaged over the whole post-shock layer; see \hyperlink{CO15}{CO15}.
      \item $^\ddagger$$a$, $b$, and $c$ are the three axis lengths (from longest to shortest) of each core. Note that $11$ ``cores" identified are actually gravitationally bound ``filaments" with aspect ratio $b/a < 0.2$, which have already been removed from the core analysis.
      \item $^\P$Notation $\measuredangle[\mu,\nu]$ represents the relative angle between vectors/axes $\mu$ and $\nu$, which by definition is within the range of $0$ and $90$ degrees.
    \end{tablenotes}
  \end{threeparttable}
%\end{center}
\end{table*}

Table~\ref{CoreSum} summarizes the physical properties measured from simulated cores. Below we discuss the orientation (Section~\ref{sec::orientation}), geometry (Section~\ref{sec::aspR}) and the relative orientation of the magnetic field (Section~\ref{sec::Bfield}) among these cores and between different models. Kinematic features related to the core's angular momentum are discussed in Section~\ref{sec::dyn}.
For reference, we note that (1) $\hat z$ is the direction of converging flow and the ``small" dimension for the post-shock layer within which; (2) $\hat x$ is the primary direction of the magnetic field in the post-shock layer (see Figure~\ref{pslayer}); (3) the filaments within which cores form are in the $\hat x$-$\hat y$ plane (see Figure~\ref{corefind} and \hyperlink{CO15}{CO15}).  

\subsection{Core Orientation}
\label{sec::orientation}

\begin{figure*}[t]
\begin{center}
\includegraphics[width=\textwidth]{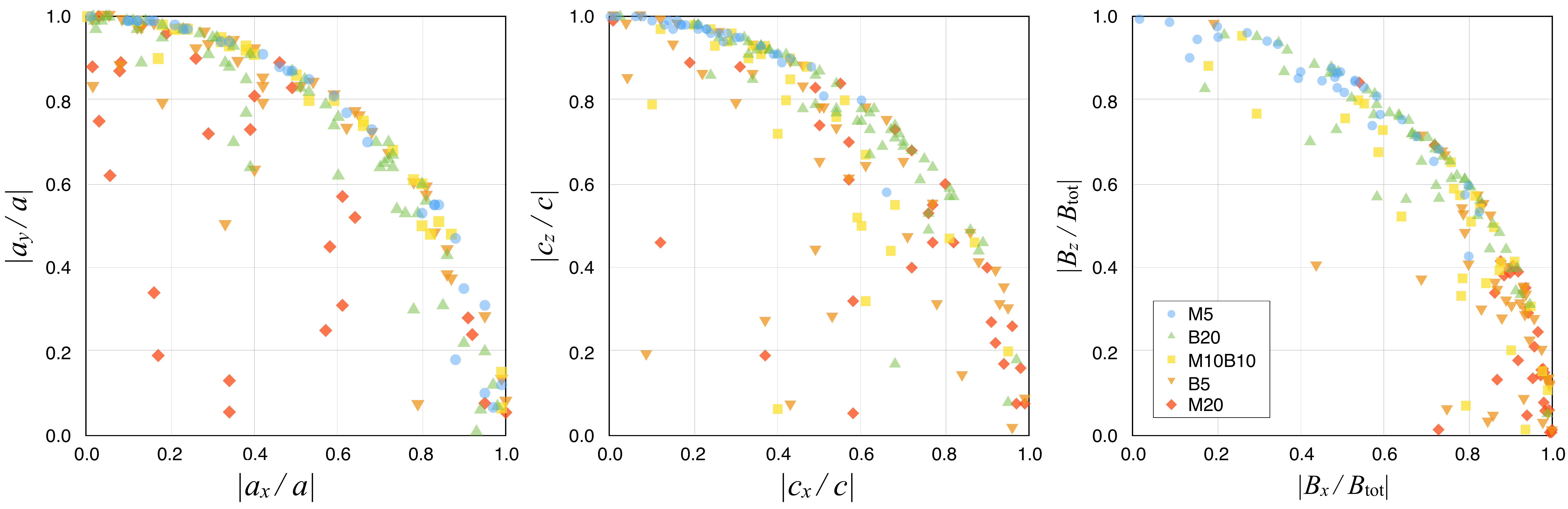}
\caption{Scatter plots of $|a_x/a|$ vs. $|a_y/a|$ (the relative $x$- and $y$-component of the major axis $a$, {\it left}), $|c_x/c|$ vs. $|c_z/c|$ (the relative $x$- and $z$- component of the minor axis $c$, {\it middle}), and $|B_x/B_\mathrm{tot}|$ vs. $|B_z/B_\mathrm{tot}|$ ({\it right}), showing that most of the cores have major axes in the $x-y$ plane, minor axes in the $x-z$ plane, and mean magnetic field in the $x-z$ plane. 
Only models with strong turbulence in the post-shock region (i.e.~without the stagnant sub-layer; models B5, M20) can form cores with large $|a_z/a|$ and $|c_y/c|$. Also note that cores formed in these models (B5 and M20) have the largest ratios $|B_x / B_\mathrm{tot}|$,
while cores formed within the stagnant sub-layers tend to have higher $|B_z / B_\mathrm{tot}|$.
}
\label{acdir}
\end{center}
\end{figure*}

For each core there are three PCA-defined axes/radii $a$, $b$, and $c$ (from longest to shortest), and we can measure their directions by calculating the relative values of their $x$-, $y$-, and $z$-components, denoted as $|a_x/a|$, $|a_y/a|$, $|a_z/a|$, and so on.
Figure~\ref{acdir} depicts the direction of the major and minor axes of all cores by showing the scatter plots of $|a_x/a|$ vs. $|a_y/a|$ (left) and $|c_x/c|$ vs. $|c_z/c|$ (middle); also shown is the scatter plot of $|B_x / B_\mathrm{tot}|$ vs. $|B_z / B_\mathrm{tot}|$, the normalized $x$- and $z$-components of the average magnetic field within individual cores.
We find that most cores have their major axes lying in the $x-y$ plane of the simulation box (${|a_x/a|}^2 + {|a_y/a|}^2 \approx 1$), with minor axes in the $x-z$ plane (${|c_x/c|}^2 + {|c_z/c|}^2 \approx 1$). Only cores formed in the more turbulent environment (models B5 and M20) can have relatively large values of $|a_z/a|$ or $|c_y/c|$ (${|a_x/a|}^2 + {|a_y/a|}^2 < 1$ or ${|c_x/c|}^2 + {|c_z/c|}^2 < 1$). 
The result that the major axis lies in the $x-y$ plane is not surprising as the converging flow along $\hat z$ creates a dense post-shock layer, within which filaments and then cores form, in the $x-y$ plane.  Especially in the case of core formation in a stagnant sub-layer, it is reasonable to expect the major axis  to be perpendicular to the inflow direction.  
Furthermore,
since it is difficult to compress the magnetic field
(which is roughly in the $x$-direction in the post-shock region) in the $y$-direction, 
minor axes are likely to lie in the $x-z$ plane.

Similarly, the right panel of Figure~\ref{acdir} shows that core-scale magnetic field tends to lie in the $x-z$ plane, with $({B_x}^2 + {B_z}^2)/{B_\mathrm{tot}}^2 \approx 1$. More importantly, there is a systematic variation of the magnetic field direction between different models, from preferably along $z$ ($|B_z / B_\mathrm{tot}| \approx 1$) for models M5 and B20, to preferably along $x$ (models B5 and M20). This result is consistent with the magnetic field direction in the post-shock region; for oblique MHD shocks propagating along $z$, the direction of post-shock magnetic field $B'$ has direction $B'_x / B'_z = r_B B_{0,x}/B_{0,z}$ relative to the pre-shock magnetic field $B_0$ \citep[see Equation~(7) in][]{2017ApJ...847...140C}, where $r_B$ is the compression ratio of $B_x$. For strong shocks, $r_B \sim v_0/v_{\mathrm{A}x0}$ (see derivations in \hyperlink{CO14}{CO14}), which means $B'_x / B'_z \propto v_0 / (B_{0,x}B_{0,z})\propto v_0 / {B_0}^2$. Models with weaker inflow Mach number (M5) or stronger pre-shock magnetic field (B20) therefore have post-shock magnetic field better aligned along $z$, while models with stronger inflow Mach number (M20) or weaker pre-shock magnetic field (B5) tend to have post-shock magnetic field almost parallel to $x$-direction.
This suggests that, regardless the presence of the stagnant sub-layer, cores barely alter the magnetic field structure before they reach the collapse stage.

\subsection{Aspect Ratio}
\label{sec::aspR}

The three PCA-defined axis lengths $a$, $b$, and $c$ yield three distinct aspect ratios, which provide a measurement of how spherical a core is. The ratio $c/b$ indicates whether a core is more prolate ($c/b \sim 1$), while the ratio $b/a \sim 1$ indicates an oblate core.  Cores with $c/b < 1$ and $b/a<1$ are triaxial. The median values of these ratios of each simulation model are listed in Table~\ref{CoreSum}. Note that ``cores" with $b/a < 0.2$ (i.e.~the major axis is at least 5 times longer than the other two axes) are considered elongated filaments and are not included in this study.

\begin{figure}
\begin{center}
\includegraphics[width=\columnwidth]{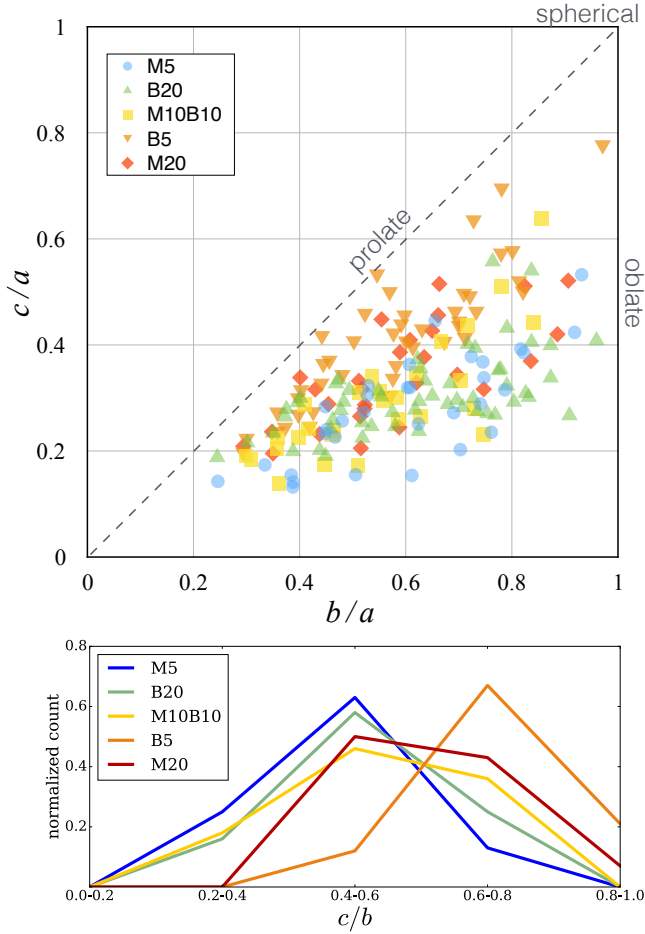}
\caption{The $c/a$ vs. $b/a$ scatter plot ({\it top}) and the histogram of $c/b$ ({\it bottom}) of dense cores formed from various simulation models.
Cores are in general triaxial, and somewhat closer to prolate than oblate.
Models with strong turbulence/weak magnetization (M20 and B5) seem to have more prolate cores ($c \sim b$), and models with weak turbulence/strong magnetization (M5 and B20) seem to have more triaxial cores ($b > c$).}
\label{aspRatio}
\end{center}
\end{figure}

Figure~\ref{aspRatio} shows a scatter plot ($c/a$ vs. $b/a$; top panel) and a histogram ($c/b$; bottom panel) of the ratios between the three axis lengths of simulated cores. Looking at the top panel of Figure~\ref{aspRatio}, these cores have a wide range of axis ratios, and in general fall within the ``triaxial" regime. Among cores from different models, models with stronger turbulence (M20) or weaker magnetization (B5) seem to have more prolate cores (closer to the diagonal $b=c$), while models with weaker turbulence (M5) or stronger magnetization (B20) tend to have low $c/a$ ($< 0.5$) and are more triaxial. 
This tendency can also be clearly observed in the bottom panel of Figure~\ref{aspRatio}, where the median value of $c/b$ shifts from $\sim 0.5$ for models M5 and B20, to $\sim 0.6$ for models M10B10 and M20, to $\sim 0.7$ for model B5.

We argue that this difference in aspect ratios of dense cores is caused by the existence of a stagnant sub-layer in some environments. As we have shown in Figure~\ref{acdir}, cores formed within the stagnant sub-layers (models M5 and B20) tend to have their minor axes $c$ along $\hat z$, because the stagnant sub-layers put limits on the core growth along the $z$ direction by its thickness (see Figure~\ref{pslayer}). In this situation core formation is roughly two-dimensional in the $x-y$ plane, within the sub-layer.
On the other hand, in the situation without the presence of a stagnant sub-layer, local turbulence and the post-shock gas flow (which roughly follows the post-shock magnetic field direction on the $x-z$ plane; see discussion in \citealt{2017ApJ...847...140C}) 
will lead to a more perturbed core-forming process
even though the magnetic pressure is dominant in the post-shock region.
We note that our simulated cores typically have $b/a \sim 0.6$ and $c/a \sim 0.3$ (see Table~\ref{CoreSum}), inconsistent with the classical picture, 
in which cores are expected to be oblate ($a\sim b\gg c$) in strongly-magnetized regions (like the post-shock regions of our simulations).

\begin{figure*}
\begin{center}
\includegraphics[width=\textwidth]{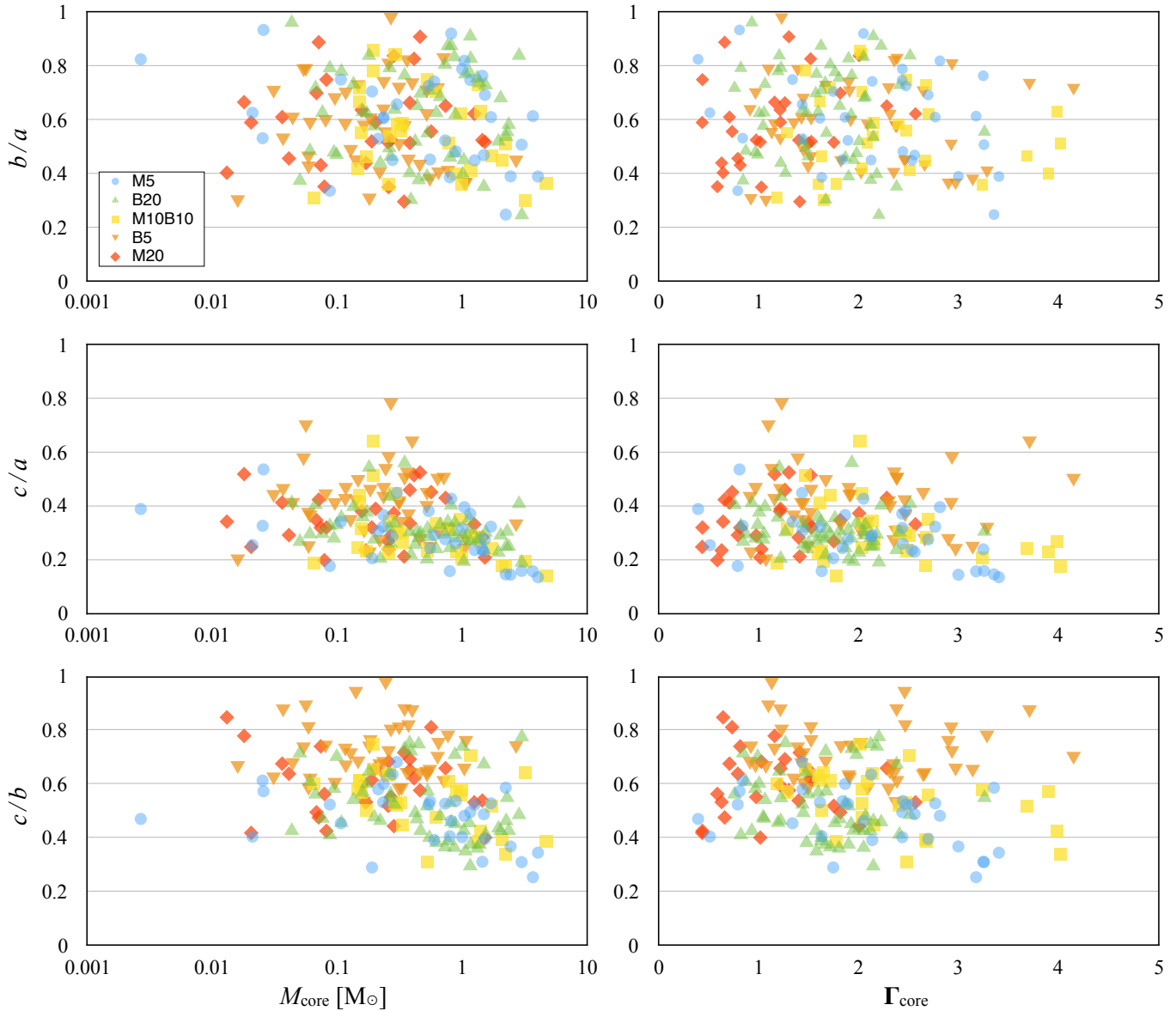}
\caption{The three aspect ratios $b/a$ ({\it top panels}), $c/a$ ({\it middle panels}), and $c/b$ ({\it bottom panels}) as functions of the core mass ({\it left}) and normalized mass-to-magnetic flux ratio $\Gamma$ ({\it right}), for all cores formed in various simulation models. 
The core mass is somewhat correlated with core shape, in that lower-mass cores tend to be more prolate ($c\sim b$) than higher-mass cores.
On the other hand, 
magnetization (as measured by the mass-to-flux ratio relative to the critical value) seems to have little correlation with core geometry, and in particular there is no indication that strongly-magnetized cores (low $\Gamma$) tend to be oblate with $b\sim a$, as would be true in the classical picture of star formation.
}
\label{aspRatioMassGamma}
\end{center}
\end{figure*}

Interestingly, the aspect ratio of dense cores does not seem to be strongly affected by the magnetization of the cores themselves. Figure~\ref{aspRatioMassGamma} shows the scatter plots of the three aspect ratios $b/a$ ({top panels}), $c/a$ ({middle panels}), and $c/b$ ({bottom panels}) as functions of core mass ({left panels}) and normalized mass-to-magnetic flux ratio $\Gamma$ ({right panels}). The distributions of all three aspect ratios seem totally random with respect to $\Gamma$. 
This suggests that the magnetization level within cores is not a dominant factor controlling core geometry, at least at the evolutionary stage we measure.
In contrast, 
there is a tendency for more massive cores to have smaller $c/a$ and $c/b$, and for less massive cores to be more prolate ($c\sim b$).
In detail, we find that while the overall core volume increases $\propto M^{3/2}$ (due to the $M \propto R^2$ correlation discussed in \hyperlink{CO15}{CO15}), the range of $a$ exceeds the range of $c$ (and $d\log a/d\log M> d\log c/d\log M$ at the high-mass end).
Therefore, this may simply reflect the fact that the most massive cores have larger $a$.

\subsection{Magnetic Field Orientation}
\label{sec::Bfield}

\begin{figure}[t]
\begin{center}
\includegraphics[width=\columnwidth]{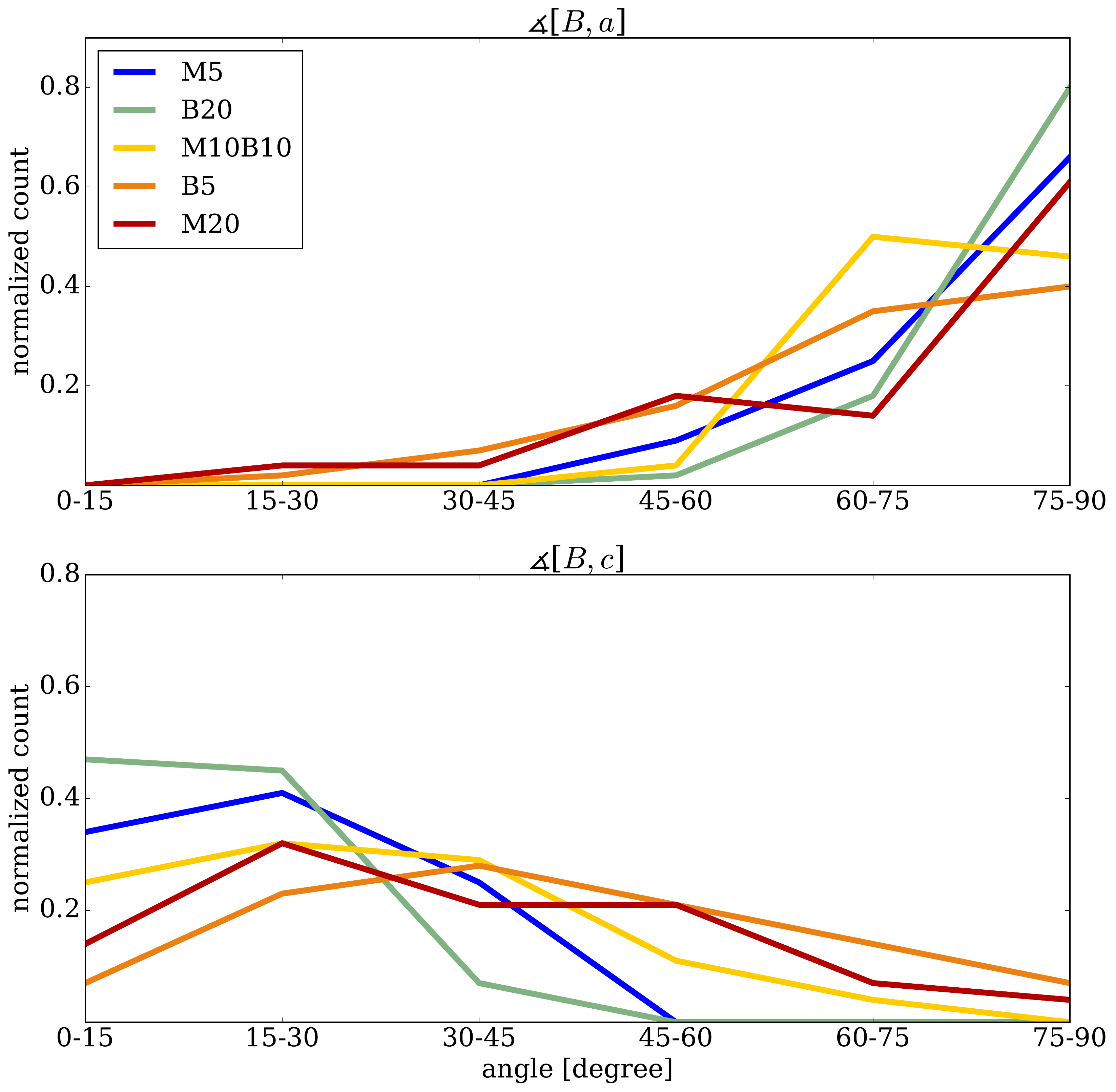}
\caption{Histogram of relative orientations between the mean magnetic field and the major ($a$, {\it top}) and minor ($c$, {\it bottom}) axes of the core. It is clear that the magnetic field preferably aligns perpendicular to the major axis and parallel to the minor axis, especially in models with stronger magnetization (B20) or weaker turbulence (M5). The alignment becomes weaker when the cloud is more perturbed (model M20) or weakly magnetized (model B5). }
\label{hist_B}
\end{center}
\end{figure}

It is interesting to investigate possible alignment between the magnetic field and core geometry.
Figure~\ref{hist_B} shows the histograms of the angle between the mean magnetic field within the core and its major (top panel) and minor (bottom panel) axes. These distributions show that cores have preferential alignments with respect to the local magnetic field, with minor axes preferentially more parallel to the field and major axes preferentially more perpendicular to the field. This is especially true in models with stronger pre-shock magnetic field (B20) or weaker pre-shock inflow Mach number (M5), which results in a less-perturbed post-shock environment, including the stagnant sub-layer, in which the cores form. 
However, we note that even though the longest axis is preferentially perpendicular to the magnetic field, cores are not strictly oblate ($a\sim b$), as would be expected for the very strongly magnetized case based on the classical picture of star formation.
Also, there is no evidence that more oblate cores have short axes better aligned with the magnetic field, as would be expected in the classical picture. 
On the other hand, only in models without the stagnant sub-layers in the post-shock regions (highly turbulent (M20) or weakly magnetized (B5) models) do we find magnetic fields almost perpendicular to the core minor axes (large $\measuredangle$[$B$, $c$]), because the strong velocity turbulence can interfere the process of anisotropic condensation along magnetic field lines.
However, we caution that cores tend to be more prolate ($b\sim c$) in these models (see Figure~\ref{aspRatio}), and therefore the direction of $c$ itself is less significant in this situation.

\section{Kinematic Features}
\label{sec::dyn}

\renewcommand{\arraystretch}{1.1}
\begin{table*}[t]
\begin{center}
  \begin{threeparttable}
\caption{Kinetic features of identified cores.$^{\dagger}$}
\label{CoreKE}
\vspace{.1in}
\begin{tabular}{ l || c c | c c c c | c c}
  \hline
  Model & \# Cores & $E_K$$^{\star}$ & $\Omega$ & $E_\mathrm{rot}$$^{\star}$ & $E_\mathrm{turb}$$^{\star}$ & $E_\mathrm{rot}/$ & $\Omega_\mathrm{ring}$ &$E_\mathrm{rot,ring}/$ \\ 
   & considered$^{\S}$ & $/10^{-3}$ & (km/s/pc) & $/10^{-4}$ & $/10^{-3}$ & $E_\mathrm{total}$$^{\ddagger}$ & (km/s/pc) & $E_\mathrm{total,ring}$$^{\ddagger}$   \\
  \hline
  M5 & 29  & 7.92 & 2.17  & 2.97 & 7.31 & $6.4~\%$ & 2.72 & ~$7.8~\%$  \\
  B20 & 46  & 8.45 & 3.86 & 3.46 & 8.35 & $4.8~\%$ & 5.54 & $11.2~\%$ \\
  M10B10 & 25 & 8.67 & 4.13 & 4.54 & 7.91 & $5.3~\%$ & 5.23 & $11.7~\%$  \\
  B5 & 33 & 4.86 & 6.09 & 4.72 & 3.88 & $8.9~\%$ & 8.14 & $14.0~\%$  \\
  M20 & 22 & 5.47 & 8.21 & 5.07 & 4.91 & $6.8~\%$ & 15.26 & $12.1~\%$  \\
  \hline 
\end{tabular}
    \begin{tablenotes}
      \footnotesize
      \item $^\dagger$Columns (3)$-$(9) are median values over all cores for each parameter set (6 simulation runs).
      \item $^\S$From cores considered in Table~\ref{CoreSum}, only those with enough cells for ring fit (see Section~\ref{sec::anacode}) are considered here.
      \item $^{\star}$The unit for energy here is $M_\odot \cdot$~(km/s)$^2$.
      \item $^{\ddagger}$ Here, $E_\mathrm{total} \equiv E_\mathrm{rot} + E_\mathrm{turb}$, which is not necessary equal to $E_K \equiv \Delta V\sum \rho_i {v_i}^2/2$; see Equation~(\ref{eq::Eturb}).
    \end{tablenotes}
  \end{threeparttable}
\end{center}
\end{table*}

\subsection{Integrated Angular Momentum and Rotational Energy}
\label{sec::netL}

The median values of angular velocity $\Omega$ and rotational energy $E_\mathrm{rot}$ calculated from the integrated angular momentum $L$ (see Section~\ref{sec::grid}) are listed in Table~\ref{CoreKE}. Also listed are the turbulent energy $E_\mathrm{turb}$ derived using Equation~\ref{eq::Eturb}, and the resulting rotational energy ratio $E_\mathrm{rot}/E_\mathrm{total}$. In all models, $E_\mathrm{rot}/E_\mathrm{turb}\sim 0.1$, and $E_\mathrm{rot}/E_\mathrm{total}\sim 0.01-0.1$. This suggests that rotation is not the dominant motion within prestellar cores. We discuss this further in Section~\ref{sec::discussion}.

\subsubsection{The Specific Angular Momentum and Core Geometry}

\begin{figure}
\begin{center}
\includegraphics[width=\columnwidth]{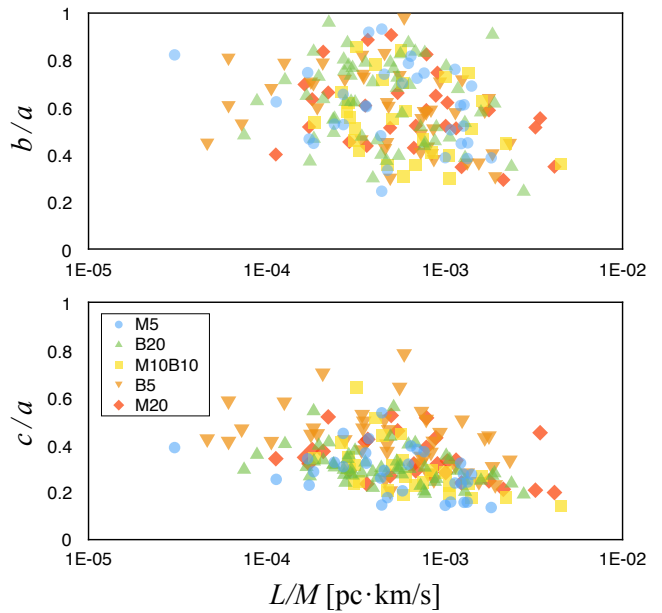}
\caption{Scatter plots of the aspect ratios $b/a$ ({\it top}) and $c/a$ ({\it bottom}) versus the specific angular momentum $L/M$. 
Cores tend to be more elongated (lower $c/a$ and $b/a$) with higher $L/M$, regardless the formation environment (different simulation models).}
\label{aspRatio_J}
\end{center}
\end{figure}

As discussed in Section~\ref{sec::grid}, the net angular momentum of a core, $L$, can be directly calculated by integrating through the core. When considering the magnitude of a dense core's angular momentum, it is more common to adopt the specific angular momentum (the angular momentum per unit mass), $L/M$, than the net angular momentum itself \citep[e.g.][]{1993ApJ...406..528G}. 
Figure~\ref{aspRatio_J} shows the scatter plots of the aspect ratios $b/a$ (top panel) and $c/a$ (bottom panel) versus the specific angular momentum. In general, both aspect ratios decrease with increasing $L/M$; this can be interpreted as faster-rotating cores being more elongated. This is opposite to the naive expectation in which faster-rotating cores are more flattened/oblate ($c/a \ll 1$, $b/a \sim 1$), similar to the case for more rapidly rotating stars or planets.
However, if we consider the definition of the net, integrated angular momentum (see Equation~(\ref{eq::netL})), Figure~\ref{aspRatio_J} may simply reflect the fact that $\Delta L_i \propto (r_i - r_\mathrm{CM})$ (for given density and velocity). More specifically, for cores with similar mass, volume, and angular velocity, those with shapes more prolate will have larger angular momentum.
This is one example of the potential risks of using the integrated angular momentum as the measurement of core rotation, because this value can be easily affected by the core geometry.

\subsubsection{Relative Orientation of the Angular Momentum}

\begin{figure}
\begin{center}
\includegraphics[width=\columnwidth]{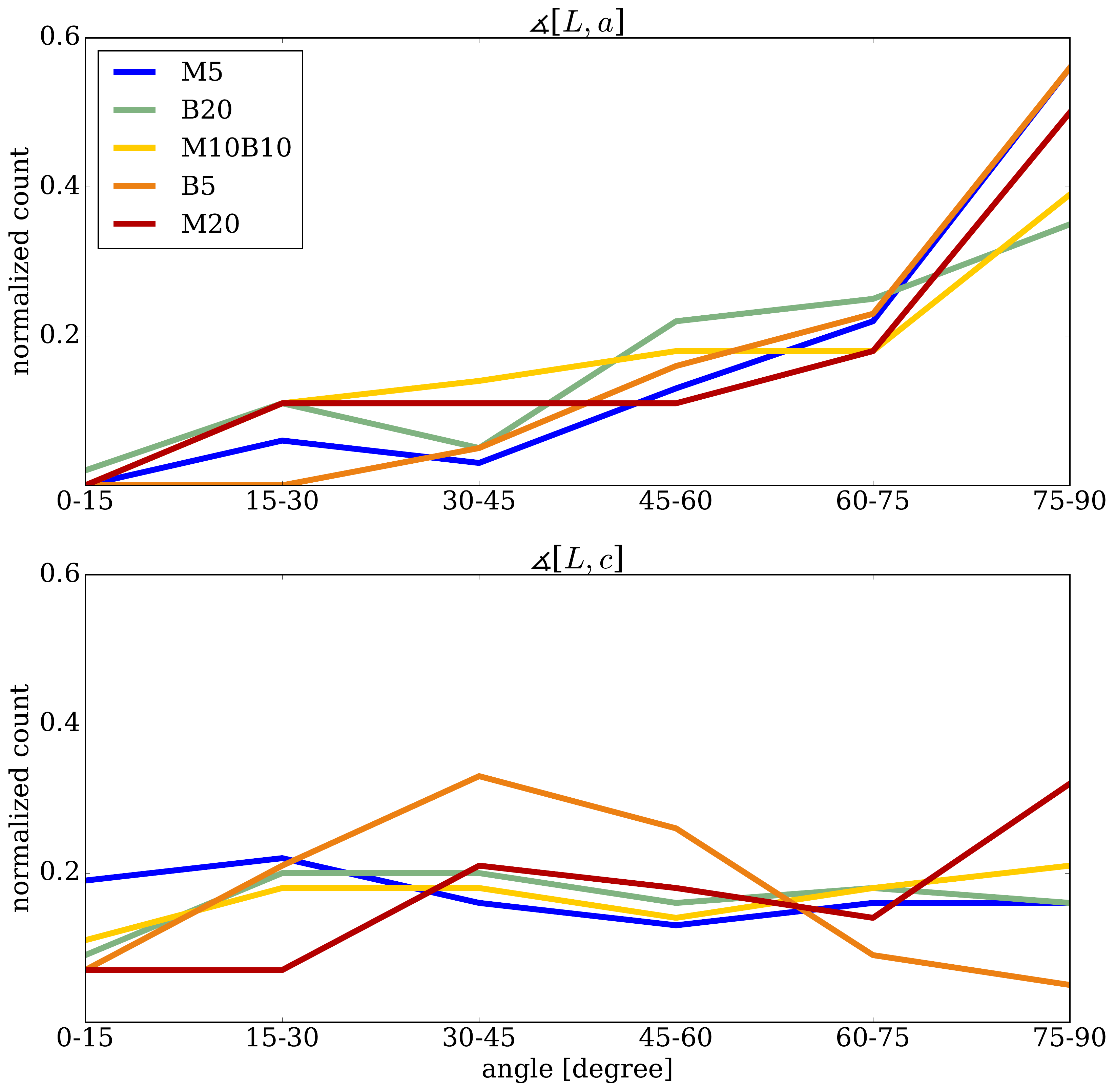}
\caption{The histogram of the relative angle between the net, integrated angular momentum ${\bf L}$ of the core and its major ({\it top}) and minor ({\it bottom}) axes. The rotational axis defined by $\hat {\bf L}$ tends to align perpendicular to the major axis, regardless the simulation models, but has no preferred direction with respect to the minor axis. }
\label{hist_L}
\end{center}
\end{figure}

Similar to Figure~\ref{hist_B}, Figure~\ref{hist_L} shows the histogram of the relative alignment of the integrated angular momentum $\mathbf{L}$ with respect to the core's major ({\it top}) and minor ({\it bottom}) axes. Though the core's net angular momentum direction appears to be perpendicular to its major axis $a$, it seems to have no correlation with the minor axis $c$. 
This again is inconsistent with the naive expectation for a rotating oblate spheroid.
We discuss this further in Section~\ref{sec::turborigin}.

\subsubsection{Rotation-Magnetic Field Misalignment}

\begin{figure}[t]
\begin{center}
\includegraphics[width=\columnwidth]{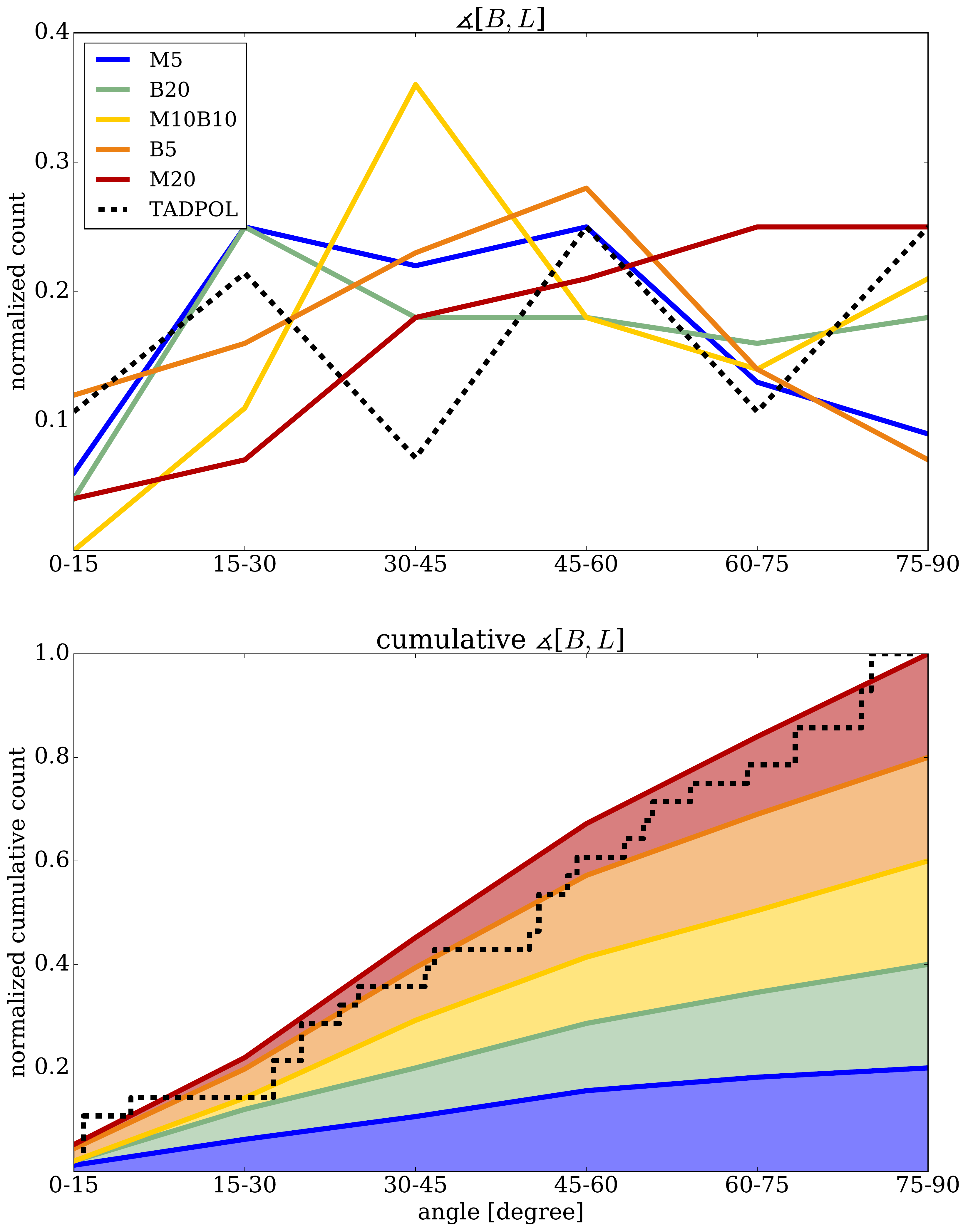}
\caption{{\it Top}: the histograms of the relative angle between the mean magnetic field $B$ and the integrated angular momentum ${\bf L}$, for all cores formed in different models. Note that the median values of $\measuredangle$[$B$, $L$] shifts from small to large angles for model sets M5, M10B10, M20 (increasing turbulence strength) and B20, M10B10, B5 (decreasing magnetization level).
{\it Bottom}: the cumulative distribution function (CDF) of $\measuredangle$[$B$, $L$] among all models, which suggests a random distribution (straight-line CDF), or that the core's rotational axis is in general misaligned with the mean magnetic field. The result from the TADPOL survey \citep{2013ApJ...768..159H,2014ApJS..213...13H} is overplotted in both panels ({\it dashed line}); note that it is presented as a step function in the bottom panel, which is purely for illustration purpose and not for quantitative comparison, as the $x$-axis scale should not be the same for histogram and step function.
}
\label{hist_BL}
\end{center}
\end{figure}

In classical theory, the rotational axis is expected to be aligned with the magnetic axis of cores because magnetic braking is faster in perpendicular compared to parallel configurations \citep{1979ApJ...228..159M,1979ApJ...230..204M}.
One of the most important breakthroughs in recent observations of magnetic field morphology within dense cores is that the magnetic field may not be aligned with the rotational axis as in classical theory \citep{2013ApJ...768..159H,2014ApJS..213...13H}. Our simulated cores provide the proper database for examining the rotation-magnetic field alignment in prestellar cores that are formed from a turbulent medium.
Figure~\ref{hist_BL} illustrates the histogram (or the probability distribution function, PDF, {\it top}) and the cumulative distribution function (CDF; {\it bottom}) of the relative angles between $\mathbf{B}$ and $\mathbf{L}$ for all cores formed in our simulations. Though the PDF of each model show variations across [$0^\circ$, $90^\circ$] and peak at different angles for different models, the CDF of all cores from various models combined appears to be a relatively straight line (random distribution) between [$0^\circ$, $90^\circ$]. This agrees with the TADPOL result \citep{2013ApJ...768..159H,2014ApJS..213...13H}, which is shown in the bottom panel of Figure~\ref{hist_BL} as a step function, and suggests that there is no preferred orientation of core's angular momentum with respect to its magnetic field. Though this picture differs from the classical model in which the rotational axis is aligned with the magnetic field, the rotation-magnetic field misalignment is critical in solving the magnetic braking catastrophe in protostellar disk formation \citep[see review in][]{2014prpl.conf..173L}.

The individual PDF of $\measuredangle$[$B$, $L$] for each model (Figure~\ref{hist_BL}, {\it top}) shows that the relative angle between a core's magnetic field and integrated angular momentum depends on the turbulence strength and magnetization level of the background cloud where the core forms. By looking at models with increasing pre-shock inflow Mach numbers (M5, M10, M20), we see that the median value of $\measuredangle$[$B$, $L$] shifts from small to large angles, which is also obvious for models with decreasing pre-shock magnetic field magnitude (B20, B10, B5). This is correlated with the turbulence level in the post-shock medium where the cores are formed, which is determined by the existence of the sub-layer; if the secondary convergent flow in the post-shock region is strong enough to create the sub-layer (as the cases of models B20 and M5), cores will form within this stagnant, dense sheet where magnetic field plays a more significant role than turbulence. This relatively quiescent core forming process is similar to the classic model, which results in nearly-aligned magnetic field and angular momentum within cores. 
On the other hand, in models B5 and M20 there is no sub-layer where the post-shock gas flow velocities ($v^\prime \sim 0.1-0.2$~km/s; see Equation~(10) in \citealt{2017ApJ...847...140C}) collide and cancel before dense cores form, and therefore the post-shock gas momentum is applied directly to dense clumps as turbulence during the core forming process. 
Since the post-shock regions in these models are sub-Alfv{\'e}nic, the gas flows near forming cores are likely along the magnetic field lines; if these anisotropic gas flows are the main sources of core's angular momentum, it is not surprising that $\measuredangle$[$B$, $L$] is relatively large in these models.

\begin{figure}
\begin{center}
\includegraphics[width=\columnwidth]{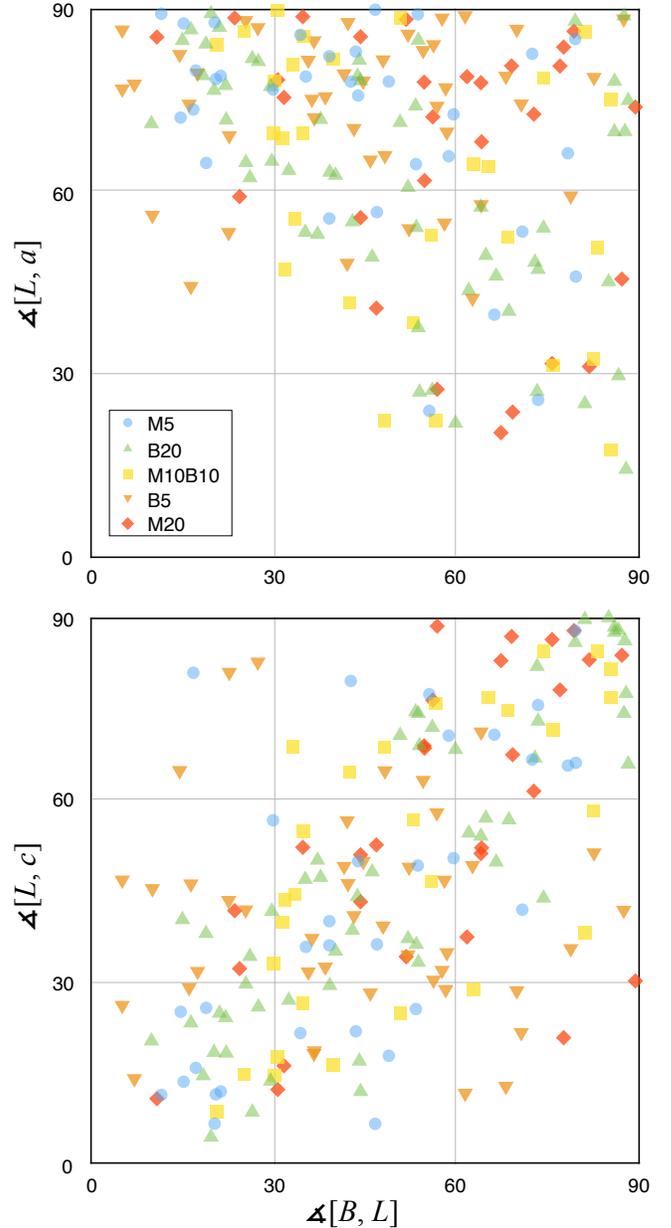}
\caption{Scatter plots of relative angle between the net angular momentum $L$ and major ({\it top})/minor ({\it bottom}) axes and the rotation-magnetic field relative angle. 
The results shown are consistent with a correlation between the magnetic field direction and the minor axis (see Figure~\ref{hist_B}).
}
\label{Lax0Lax2BL}
\end{center}
\end{figure}

Figure~\ref{Lax0Lax2BL} compares the relative angles between $L$ and $a$, $c$ with the relative angles between $B$ and $L$. Interestingly, we find that the distribution of $\measuredangle$[$L$,~$a$] vs.~$\measuredangle$[$B$, $L$] is concentrated in the upper-right half of the plot, meaning that the angles $\measuredangle$[$L$,~$a$] and $\measuredangle$[$B$, $L$] cannot both be small ($\lesssim 45^\circ$). Correspondingly, the distribution of  $\measuredangle$[$L$,~$c$] vs.~$\measuredangle$[$B$, $L$] appears to be a diagonal band from lower-left to upper-right of the plot, which means that these two angles might be positively correlated. These results agree with Figure~\ref{hist_B}, showing that the average magnetic fields within dense cores tend to align with the minor axes.

\subsection{Non-rigid-body Rotation}
\label{sec::nrbRot}

Table~\ref{CoreKE} lists the average angular velocity $\Omega_\mathrm{ring}$ and the relative rotational energy $E_\mathrm{rot,ring}/E_\mathrm{total, ring}$ derived from our ring-fit method described in Section~\ref{sec::anacode}. 
Strikingly but not surprisingly, the rotational energy ratio from our ring-fit method is around a factor of 2 larger than that measured under the assumption of rigid-body rotation. 
Though this does not guarantee that the ring-fit method is more accurate for measuring core's angular momentum (see the discussions in the Appendix),
the fact that $E_\mathrm{rot}/E_\mathrm{total}$ is only $\sim 10~\%$ in either fitting method provides strong evidence that rotation is not the dominant motion within prestellar cores.
We provide further discussion below in Section~\ref{sec::discussion}.

\section{Discussion: The Origin of Core Angular Momentum}
\label{sec::discussion}
\label{sec::turborigin}

It has been known that the specific angular momentum of observed dense cores/clumps are correlated with their sizes, approximately following a power law, $L/M \propto R^{\alpha}$, with $\alpha \approx 1.5$. This was first reported in \citet{1993ApJ...406..528G}, and was later confirmed by many follow-up studies \citep[e.g.][]{2002ApJ...572..238C,2003A&A...405..639P,2007ApJ...669.1058C}. Also, these observations suggest that the ratio between rotational energy and gravitational energy, $\beta_E \equiv (L^2/(2I))/(qGM^2/R)$ (where $q=3/5$ for a uniform density sphere; see definition in \citet{1993ApJ...406..528G}), is relatively independent of core/clump size.

\begin{figure}
\begin{center}
\includegraphics[width=\columnwidth]{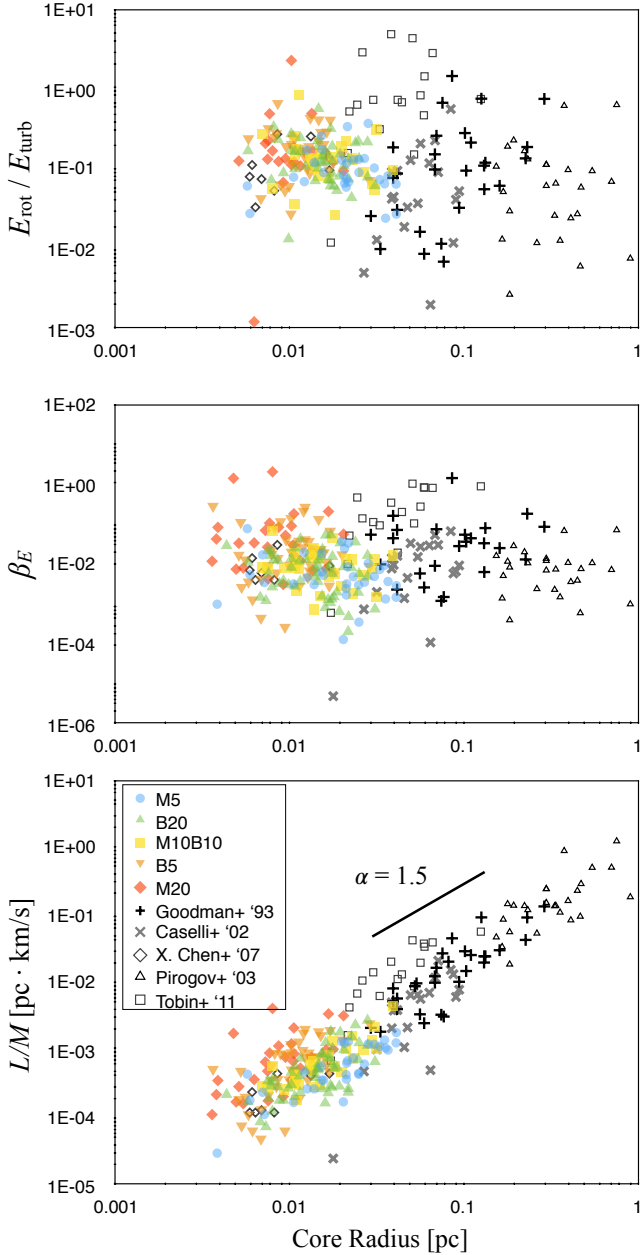}
\caption{The rotational-to-total kinetic energy ratio ({\it top}), rotational-to-gravitational energy ratio $\beta_E$ ({\it middle}) and the specific angular momentum $L/M$ ({\it bottom}) as functions of core/clump radius, from both observations (see text) and the simulated cores discussed in this study. The energy ratio distributions are independent of core size, while the specific angular momentum appears to roughly follow a power law of core size, $L/M \propto R^\alpha$ for $\alpha \sim 1.5$, over more than two orders of magnitude in spatial scales. This may suggest that the prestellar core acquires angular momentum from a much larger scale than the immediate surrounding of the core, or the so-called rotation within dense cores is inherited from turbulent motions at cloud scales.}
\label{rotEng}
\end{center}
\end{figure}

In Figure~\ref{rotEng} we show the $L/M$ vs.~$R$ relationship (bottom panel) from several observational studies \citep{1993ApJ...406..528G,2002ApJ...572..238C,2003A&A...405..639P,2007ApJ...669.1058C,2011ApJ...740...45T}. We compare to the $L/M$ vs.~$R$ relationship in our simulated cores considering $R \equiv \sqrt[3]{a\cdot b\cdot c}$ (i.e.~the geometric mean of the three axes), which follows the same correlation as the observations. 
The agreement of the simulations with observations confirms the positive correlation between the specific angular momentum and spatial scale of dense cores/clumps. 
Also plotted is the rotational-to-gravitational energy ratio $\beta_E$ (middle panel) as a function of core radius $R$ for both our simulations and observations. The independence of core rotational-to-gravitational energy ratio with core size is also confirmed.
Quantitatively, the range of $\beta_E$ for our simulated cores agrees with observations.

The fact that $L/M \sim R \cdot v_\mathrm{rot} \propto R^{3/2}$ suggests that $v_\mathrm{rot} \propto R^{1/2}$.  In combination with the well-known result that turbulent velocities increase roughly $\propto R^{1/2}$ in supersonic turbulence (both in observations and simulations; see review in \citealt{2007ARA&A..45..565M}), this suggests that the rotational velocity in cores is inherited from the overall turbulent cascade.  
In addition, the top panel of Figure~\ref{rotEng} shows the rotational-to-kinetic energy ratio, $E_\mathrm{rot} / E_K$, as a function of core radius $R$ for both simulated and observed cores.\footnote{To estimate the kinetic energy of observed cores, we used the total observed linewidths $\sigma_v$ within cores reported by the cited observation studies and calculated $E_K \approx 3/2 \cdot M_\mathrm{core} {\sigma_v}^2$.} For both simulated and observed cores, the range of $E_\mathrm{rot} / E_K$ is similar and independent of $R$. This suggests that whatever size a core/clump is, it is sampling from the turbulence at that corresponding scale in setting its rotation.  
Though this rotational energy could be sub-dominant at core scales, it is essential to subsequent disk formation. Once a star-disk system forms in the interior (at much smaller scales than that studied here), the core's velocity structure (and angular momentum) may be altered by outflows and/or other feedback mechanisms \citep[see e.g.][]{2017ApJ...847..104O}.

To investigate the accuracy of our estimated core angular momentum, we ran a set of tests to examine the analysis method, which is described in the Appendix. These tests show that the measured $E_\mathrm{rot}/E_K$ ratio within the core could in principle reflect the relative significance of turbulence with respect to rigid-body rotation (see Figure~\ref{ErotEturb}, right panels). More importantly, we showed that for a pure-turbulent core (net angular velocity $\Omega=0$), the ``projection" of turbulence within it will naturally lead to $E_\mathrm{rot}/E_K \sim 0.1$ (Figure~\ref{ErotEturb}, left panels), consistent with the values measured from our simulated cores.
We therefore conclude that rotation is not the dominant motion within prestellar cores, and the ratio between the turbulence amplitude $\sigma_v$ and maximum rotational speed ($v_\mathrm{rot,max} \sim \Omega \cdot R_\mathrm{core}$) must be $\gtrsim 1$ within prestellar cores.

\section{Summary}
\label{sec::summary}

In this paper, we investigated the $> 100$ dense cores formed naturally in the \hyperlink{CO15}{CO15} MHD simulations to examine the structural, magnetic, and kinetic properties of prestellar cores with masses $M_\mathrm{core}\sim 0.01-5~\mathrm{M_\odot}$ and sizes $R_\mathrm{core}\sim 0.005-0.1$~pc.
We found that our simulated cores are generally triaxial, unlike the idealized oblate cores of classical theory.
We showed that environmental effect plays an important role in shaping prestellar cores, especially by providing spatial constraints via ram or magnetic pressure.
In addition, the formation of prestellar cores is strongly affected by gas turbulence, in the way that cores acquire rotational energy from local turbulence, which leads to the misalignment between magnetic fields and rotational axes within dense cores.

Our main conclusions are as follows: 
\begin{enumerate}

\item
When present, a stagnant sub-layer \citep[see discussions in][]{2017ApJ...847...140C} in the post-shock region (Figure~\ref{pslayer}) is critical in setting up the environment wherein prestellar cores form. Core formation within this sub-layer (models M5 and B20) is more quiescent and more similar to classical theory, while cores formed without this sub-layer (models M20 and B5) are more disturbed by local gas turbulence.

\item
Cores preferentially have their major axes in the plane parallel to the shock front ($x$-$y$ plane; see Figure~\ref{acdir}, left), because the ram pressure of inflow limits core growth along $z$.
This might help explain the mass-size relation of dense cores reported in both numerical (\hyperlink{CO15}{CO15}) and observational \citep{2013MNRAS.432.1424K} studies, $M\propto R^k$ with $k\sim 2-2.5$, because for cores formed within shock-compressed, locally-flat regions core growth is basically two-dimensional.
On the other hand, we find that most of the cores have both their minor axes and mean magnetic fields lying in the $x$-$z$ plane (Figure~\ref{acdir}, middle and right panels) defined by the direction of inflow and the cloud-scale magnetic field. The minor axis is rarely along the $\hat y$ direction because it is difficult for the gas to flow or compress the magnetic field in this direction.

\item
Though cores are generally triaxial (Figure~\ref{aspRatio}, top) rather than having the oblate shape often adopted in classical theory, the core-scale magnetic field is still generally aligned with core's minor axis and perpendicular to the major axis (Figure~\ref{hist_B}). Only those cores formed under a more perturbed environment (without the stagnant sub-layer; models B5 and M20) can have magnetic field nearly perpendicular to their minor axes. However, these cores also tend to be more prolate ($b\sim c$; see Figure~\ref{aspRatio}, bottom), which means the direction of $c$ is less meaningful in these cases.

\item
The integrated angular momentum vector within cores does not have a preferred orientation with respect to the core's three axes (Figure~\ref{hist_L}), except being generally perpendicular to the major axis, which is a mathematical result from the definition of angular momentum (for cell $i$ at a distance of $r_i$ from the rotational axis, its angular momentum $L_i \propto r_i$). More importantly, there is no preferred alignment between the magnetic field and angular momentum within cores (Figure~\ref{hist_BL}), as reported in the TADPOL observational survey \citep{2014ApJS..213...13H} and a follow-up numerical study considering different viewing angles of two simulated protostellar envelopes \citep{2017ApJ...834..201L}. Since misalignment between a core's rotational axis and magnetic field may be critical in reducing magnetic braking during core collapse, this may be important to understanding of protostellar disk formation.

\item
Our analyses indicate that the commonly-adopted assumption of rigid-body rotation may underestimate the rotational motion in most dense cores. We presented a new method of calculating core angular momentum, {\it the ring-fit} (see Sections~\ref{sec::anacode} and \ref{sec::nrbRot}), which gives a factor of 2 higher measurement of rotational energy (see Table~\ref{CoreKE}). Our results also suggest that the measured angular momentum within cores could simply be from the projection of ambient turbulence at core scale (see Figure~\ref{ErotEturb}) as previously suggested by \cite{2000ApJ...543..822B}.

\item
With our detailed analysis of core-scale kinematics, we have revisited the specific angular momentum$-$size correlation of dense cores/clumps reported in many observations (Figure~\ref{rotEng}, bottom). Our simulated cores fit with the observational results well, and extend to smaller spatial scales. 
The correlation of $L/M$ with $R$ over two orders of magnitude in spatial scale suggests that ``rotation" within these cores/clumps shares the same origin with the velocity scaling consistent with larger-scale turbulence.
We find that the rotational-to-gravitational energy ratio $\beta_E$ (Figure~\ref{rotEng}, middle) and the relative rotational energy $E_\mathrm{rot}/E_K$ ((Figure~\ref{rotEng}, top) have similar ranges and are independent of scale for both simulated and observed cores.
Taken together, these results suggest that prestellar cores inherit their original angular momentum from cloud-scale turbulence, which may in part be driven by feedback (outflows, etc.) from other stars that formed earlier.  

\end{enumerate}

%a paragraph added to the conclusions section saying a bit more about various simulation idealizations and their potential effects:
%? missing feedback, and analysis of cores at t <= singularity in first core implies we are only considering rotation in early (pre stellar) evolutionary stages
%? idealized converging flow without large scales implies potential effects on core rotation of larger scale turbulence (including turbulence driven by feedback in other nearby stars) is missing 
%? assumption of uniform magnetic fields in initial conditions; lack of magnetic field variation may affect turbulence and hence rotation
%? limited parameter range studied in terms of B-field, density, converging velocity, etc.  

We note that our simulations and analyses only focus on cores in early (prestellar) evolutionary stages and therefore do not include effects from stellar feedback. There are also various idealizations in our simulations that could potentially affect our conclusions. The converging-flow setup intrinsically excludes  scales $\gtrsim L_\mathrm{box}$, and therefore cannot capture effects of large-scale turbulence (including turbulence driven by feedback in other nearby stars) in development of core rotation.  
Also, we assumed uniform magnetic fields in the initial conditions, and this lack of magnetic field variation may affect the structure of local turbulence and hence rotation at core scales. Our results could also be biased by the limited parameter range (in terms of magnetic field strength, gas density, inflow velocity, etc.) investigated in this study. Nevertheless, the connection between core-scale angular momentum and the immediately-surrounding cloud-scale turbulence is clear in both our numerical results and previous observations.

%{\bf We note that our analyses only focus on cores in early (prestellar) evolutionary stages and does not include the effect from feedback. There are also various idealizations in our simulations that could potentially affect our conclusions. The converging-flow setup intrinsically excludes turbulence at scales $\gtrsim L_\mathrm{box}$, and therefore core rotation from larger-scale turbulence (including turbulence driven by feedback in other nearby stars) is missing. Also, we assumed uniform magnetic fields in the initial conditions, and this lack of magnetic field variation may affect the structure of local turbulence and hence rotation at core scales. Our results could also be biased by the limited parameter range (in terms of magnetic field strength, gas density, inflow velocity, etc.) investigated in this study. Nevertheless, the connection between the original core-scale angular momentum and the cloud-scale turbulence is clear in both our numerical results and previous observations.}

\acknowledgements
We thank the referee for a very helpful report.
C.-Y.C. is grateful for the support from Virginia Institute of Theoretical Astronomy (VITA) at the University of Virginia through the VITA Postdoctoral Prize Fellowship, and the support partly from NSF grant AST-1815784.
The work of E.C.O. was supported by grant 510940 from the Simons Foundation.  

\appendix
\section{The Significance of Integrated and Ring-fitted Angular Momenta}

\begin{figure*}
\begin{center}
\includegraphics[width=\textwidth]{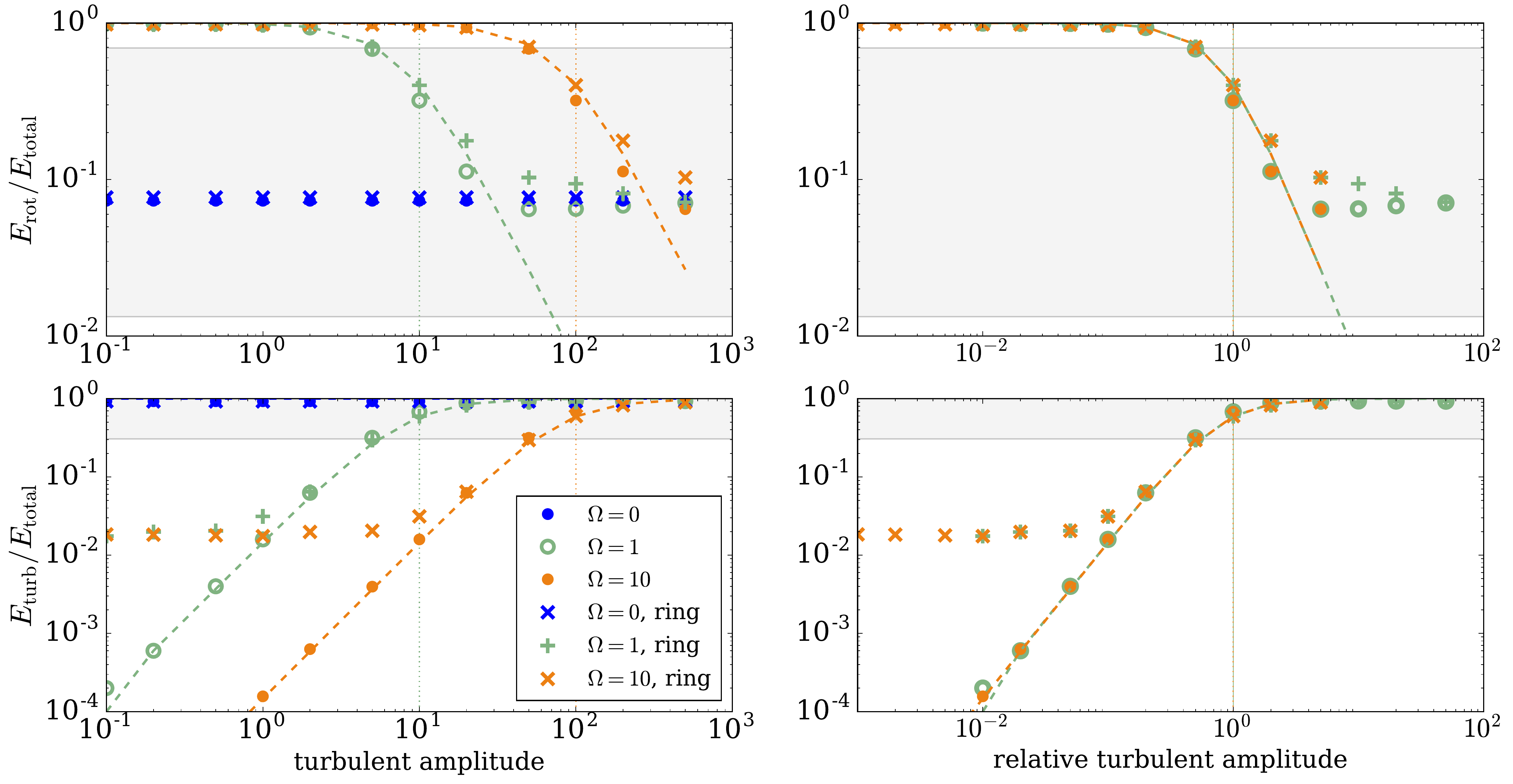}
\caption{The ratios between rotational energy $E_\mathrm{rot}$ ({\it top panels}) and turbulent energy $E_\mathrm{turb}$ ({\it bottom panels}) to the total kinetic energy $E_\mathrm{total}$ inside dense cores. The test cases are turbulent ellipsoidal cores with rigid-body angular speed $\Omega=0$ ({\it blue dot}), $1$ ({\it green diamonds}), and $10$ ({\it orange crosses}). {\it Left panels}: energy ratios as functions of the turbulent amplitude ($\sigma_v$) within the core. {\it Right panels}: energy ratios as functions of the relative turbulent amplitude ($\sigma_v / v_\mathrm{rot, max}$) within the core. For each test model, the theoretical values of $E_\mathrm{rot}/E_\mathrm{total}$ and $E_\mathrm{turb}/E_\mathrm{total}$ are also shown ({\it dashed lines}) as references of the accuracy of the fitted values.
The ranges of $E_\mathrm{rot}/E_\mathrm{total}$ and $E_\mathrm{turb}/E_\mathrm{total}$ measured from simulated prestellar cores (using the ring-fit method) are also overplotted ({\it grey bands}).}
\label{ErotEturb}
\end{center}
\end{figure*}

Here we describe the numerical tests that we conducted to examine our method of measuring angular momentum, for both the traditional rigid-body fit and our newly-developed ring fit.
We constructed an ellipsoid with uniform density, $\rho=1$ in code units, and size $a =20$ and $b = c = 10$ cells. This ``core" is assigned a rigid-body angular speed $\Omega = 0$, $1$, or $10$ in code units along $a$, plus a perturbed random velocity field with average amplitude $\sigma_v$ ranging from $0.1$ to $500$ in code units; this leads to a range of rotational speed within the core ($v_\mathrm{rot, max} = 10$ for $\Omega=1$ and $100$ for $\Omega=10$). We then applied the same analysis for cores identified in our full simulations and measure the angular momentum, rotational energy, and turbulent energy of the core, and compared these values directly to imposed values. 

\subsection{The Dependence of Measured Angular Momentum on Turbulent Level}
\label{sec::ErotEturb}

The direct results are shown in the left panel of Figure~\ref{ErotEturb}. The $\Omega = 0$ case gives the inherent numerical error in our measurement of $E_\mathrm{rot}/E_\mathrm{total}$, which is $\sim 10~\%$ for both rigid-body and ring-fit methods in this case (we will discuss the dependence of this numerical error on cell resolution in Section~\ref{sec::ErotEturbSmall}).
This error value represents the minimum of $E_\mathrm{rot}/E_\mathrm{total}$ ratio that can be correctly measured in this core; 
in both the $\Omega=1$ and $\Omega=10$ cases
the measured $E_\mathrm{rot}/E_\mathrm{total}$ values are truncated by the numerical error at large $\sigma_v$. 
Similarly, there is a numerical error for the $E_\mathrm{turb}/E_\mathrm{total}$ ratio ($\sim 10^{-2}$) at small $\sigma_v$, but only for the ring-fit method (see discussion below).

The results from the two test cases $\Omega=1$ and $\Omega=10$ can be combined by plotting against the {\it relative turbulent amplitude}, which is defined as $\sigma_v / v_\mathrm{rot, max}$ (right panel of Figure~\ref{ErotEturb}). 
We clearly see that both energy ratios change significantly when the turbulent level is around the same value as the maximum rotational speed ($\sigma_v / v_\mathrm{rot, max} = 1$, indicated by the vertical lines in both plots). In addition,
when the turbulent amplitude is about 5 times higher than the maximum rotational speed ($\sigma_v / v_\mathrm{rot, max} \gtrsim 5$), the measured $E_\mathrm{rot}/E_\mathrm{total}$ ratios from both fitting methods can only serve as upper limits on the imposed values.

On the other hand, when the turbulent amplitude is only about $0.1$ of the maximum rotational speed ($\sigma_v/v_\mathrm{rot,max}\lesssim 0.1$), the measured $E_\mathrm{turb}/E_\mathrm{total}$ ratios from ring fit are truncated by the numerical error and again only serve as upper limits of the real values. 
In contrast, the rigid-body fit is not affected by the same truncating value, and has fairly accurate $E_\mathrm{turb}/E_\mathrm{total}$ ratios even beyond $\sigma_v/v_\mathrm{rot,max}\lesssim 0.01$. This is likely because the velocity field within the core is constructed from rigid-body rotation, and therefore the ring-fit method will not have better performance than the rigid-body fit, especially when the turbulence amplitude is small (i.e. more comparable with pure rigid-body rotation). 

With these results, we can in principle use the $E_\mathrm{rot}/E_\mathrm{total}$ and $E_\mathrm{turb}/E_\mathrm{total}$ values measured from our simulated prestellar cores to estimate $\sigma_v / v_\mathrm{rot, max}$ in cores. Unfortunately, the range of $E_\mathrm{rot}/E_\mathrm{total}$ measured in our simulations is mostly outside the zone where we can precisely estimate the turbulent amplitude (see the shaded regions in Figure~\ref{ErotEturb}).
Nevertheless, we note that generally $\sigma_v / v_\mathrm{rot, max} \gtrsim 1$ inside prestellar cores (for both our simulated cores and for observed cores). This suggests that rotation is not the dominant motion within prestellar cores, and the measured angular momentum is indeed a projection of the turbulent velocity inside cores.

\subsection{The Dependence of Measured Angular Momentum on Cell Resolution}
\label{sec::ErotEturbSmall}

\begin{figure*}
\begin{center}
\includegraphics[width=\textwidth]{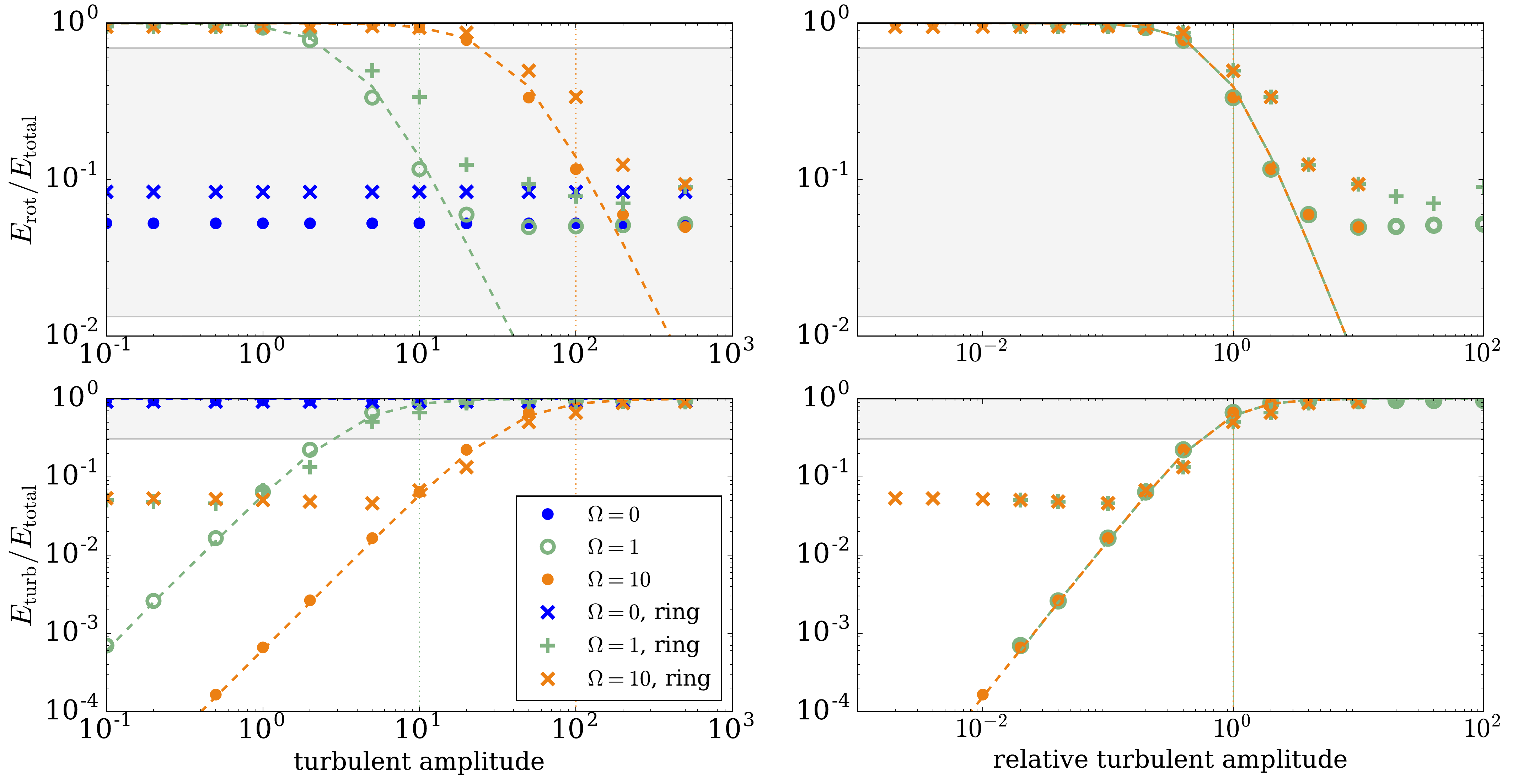}
\caption{Same as Figure~\ref{ErotEturb}, but with less cells in core.}
\label{ErotEturbSmall}
\end{center}
\end{figure*}

The numerical errors in $E_\mathrm{rot}/E_\mathrm{total}$ and $E_\mathrm{turb}/E_\mathrm{total}$ depend on the grid resolution, i.e.~number of cells inside the core. The test case described in Section~\ref{sec::ErotEturb} has $\sim 8000$ cells within the core, which is similar to some of the simulated cores considered in our main study.
We repeated the same process on cores with more ($\sim 10^6$, $a =50$ and $b = c = 40$ cells) and less ($\sim 1000$, $a =10$ and $b = c = 5$ cells; see Figure~\ref{ErotEturbSmall}) cells, and the truncating value of the $E_\mathrm{turb}/E_\mathrm{total}$ ratio measured from the ring-fit method is $\sim 10^{-3}$ and $\sim 5\times 10^{-2}$, respectively. 
Since cores formed in our simulations have numbers of cells ranging from $\sim 30$ to $\sim 40000$,
the numerical errors for those cores will therefore be 
similar to that in the test cases. Even if the error in $E_\mathrm{turb}/E_\mathrm{total}$ is slightly worse than $\sim 10^{-2}$, it will not affect the conclusions we draw from this test.

We also compared the ring-fit method with the rigid-body fit based on their performances under different resolution. From Figures~\ref{ErotEturb} and \ref{ErotEturbSmall}, we found that the accuracy of ring-fit method highly depends on the number of cells within cores, even though the $E_\mathrm{rot}/E_\mathrm{total}$ ratios measured from ring-fit method are always higher than that from rigid-body rotation. 
At higher resolution (e.g.~Figure~\ref{ErotEturb}), ring-fitted rotational energy ratio $E_\mathrm{rot}/E_\mathrm{total}$ follows the theoretical values better than the rigid-body fit; however, at lower resolution (Figure~\ref{ErotEturbSmall}), the ring-fitted energy ratios deviate further away from the theoretical values (and the deviation happens at smaller turbulent amplitude) than the rigid-body fit. 
The numerical error in $E_\mathrm{rot}/E_\mathrm{total}$ (the truncating value) from ring fit is also much larger than that in rigid-body fit for less-resolved core. Therefore, we conclude that the ring-fit is more applicable than the rigid-body fit only in larger (better resolved) cores.

\end{document}